%% file: ttspin.tex
\documentclass[aps,prd,reprint,lengthcheck,showpacs,superscriptaddress,nofootinbib]{revtex4-1}
\pdfoutput=1
\usepackage{atlasphysics}

\usepackage{preprintcover}  
\PreprintCoverPaperTitle{Measurements of spin correlation in top--antitop quark events 
from proton--proton collisions at 
${\sqrt{\bf s}}{ {\bf =} {\bf 7}} $~Te\hspace{-0.05cm}V using the ATLAS 
detector}  
\PreprintIdNumber{CERN-PH-EP-2014-116} 
\PreprintCoverAbstract{Measurements of spin correlation in top quark pair production are
presented using data collected with the ATLAS detector at the LHC with
proton--proton collisions at a center-of-mass energy of $7$~\tev,
corresponding to an integrated luminosity of $4.6$~fb$^{-1}$. 
Events are selected in final states with two charged leptons and at least 
two jets and in final states with one charged lepton and at least
four jets. Four different observables sensitive to different
properties of the top quark pair production mechanism are used 
to extract the correlation between the top and antitop quark
spins. Some of these observables are measured for the first time. 
The measurements are in good agreement with the
Standard Model prediction at next-to-leading-order accuracy. 
}  
\PreprintJournalName{PRD}  

\usepackage{subfigure}
\usepackage{mathrsfs}
\usepackage{amsmath}
\usepackage{slashed}
\usepackage{setspace}
\usepackage{footnote}
\usepackage[hyperindex,breaklinks]{hyperref} 
\usepackage{graphicx}
\usepackage{comment}
\usepackage{dcolumn}

\begin{document}

\title{Measurements of spin correlation in top--antitop quark events from
proton--proton collisions at 
${\sqrt{\bf s}}{ {\bf =} {\bf 7}} $~Te\hspace{-0.05cm}V using the ATLAS detector}

\author{The ATLAS Collaboration}
\begin{abstract}
Measurements of spin correlation in top quark pair production are
presented using data collected with the ATLAS detector at the LHC with
proton--proton collisions at a center-of-mass energy of $7$~\tev,
corresponding to an integrated luminosity of $4.6$~fb$^{-1}$. 
Events are selected in final states with two charged leptons and at least 
two jets and in final states with one charged lepton and at least
four jets. Four different observables sensitive to different
properties of the top quark pair production mechanism are used 
to extract the correlation between the top and antitop quark
spins. Some of these observables are measured for the first time. 
The measurements are in good agreement with the
Standard Model prediction at next-to-leading-order accuracy. 
\end{abstract}
\maketitle 

\newcommand{\alphalp}{\ensuremath{\alpha_\ell P}}
\newcommand{\alphaip}{\ensuremath{\alpha_i P}}
\newcommand{\alphalpcpc}{\ensuremath{\alpha_\ell P_{\mathrm{CPC}}}}
\newcommand{\alphalpcpv}{\ensuremath{\alpha_\ell P_{\mathrm{CPV}}}}
\newcommand{\ljets}{$\ell+$jets}
\newcommand{\ejets}{$e+$jets}
\newcommand{\mujets}{$\mu+$jets}
\newcommand{\emu}{$e^{\pm}\mu^{\mp}$}
\newcommand{\mm}{$\mu\mu$}
\newcommand{\elel}{$ee$}
\newcommand{\mcatnlo}   {\mbox{MC@NLO}}
\newcommand{\ppm}{$\pm$}
\newcommand\mll{\ensuremath{m_{\ell\ell}}}
\newcommand{\etmiss}{$\et^{\textrm{miss}}$}
\newcommand\htlep{\ensuremath{\HT}}
\newcommand\dbq{$\Delta\phi(\ell,b)$}
\newcommand\ddq{$\Delta\phi(\ell,d)$}

\DeclareGraphicsExtensions{}
\thispagestyle{empty}
\setcounter{page}{1}

\section{Introduction}
\label{sec:intro}
The top quark 
is the heaviest known elementary
particle. Besides the high mass, 
it has  the shortest lifetime of any quark, determined to be
$( 3.29^{+0.90}_{-0.63} ) \times 10^{-25}$~s~\cite{bib:toplifetime_d02},   
which is shorter than the time scale for 
hadronization~\cite{toplifetime_theorie}. 
This implies that
top quarks can be studied as bare quarks, i.e. quarks before
hadronization, and the spin information of
the top quark can be deduced from the angular distributions 
of its decay products. 

In the Standard Model (SM) of particle physics, top quarks are produced
at hadron colliders in pairs ($t\bar{t}$), predominantly via strong
interactions, or singly via the electroweak interactions. At the Large
Hadron Collider (LHC), which collided protons ($pp$) at a 
center-of-mass energy of 7 \tev\ in 2011, top quarks were mainly produced 
in pairs via gluon fusion. 
In the SM, $t\bar{t}$ pairs are produced essentially unpolarized at
hadron colliders~\cite{bernreutherpol}, as has been tested in recent 
measurements by the D0 collaboration~\cite{d0pol} and the ATLAS and
CMS collaborations~\cite{atlaspol,cmsspin}. Nonetheless, the correlation
of the spin orientation of the top quark and the top antiquark can be
studied, and is predicted to be
non-zero~\cite{Kuhn,Barger_Ohnemus_Phillips,Kane_Ladinsky_Yuan,Arens:1992fg,Mahlon:1995zn,Stelzer_Willenbrock,Brandenburg:1996df,Chang:1995ay,Bernreuther:1995cx,Dharmaratna_Goldstein,Mahlon:1997uc,Beneke:2000hk,Bernreuther:2001rq,Bernreuther:2004jv,Nelson:2005jp,Godbole:2006tq,bernreuther_top_review,bernreutherpol,bernreuther_si_chromo}.     

New physics models beyond the SM (BSM) can alter the spin correlation
of the top quark and top antiquark by modifying the production mechanism
of the $t \bar t$ pair. Also, they can modify the $t \bar t$ decay by
which the spin information is accessed. The first scenario occurs,
for example, in BSM models where a $t \bar t$ pair is produced via a
high-mass $Z^{'} $ boson~\cite{Zprime,moreZprime} or via a heavy Higgs boson that
decays into $t \bar t$~\cite{Higgs}. The second scenario occurs, for example,
in supersymmetric models if a top quark decays into a spin~zero
particle like a charged Higgs
boson, which  then decays into a lepton and a
neutrino~\cite{Hplus,Hplus_tauonic}. Thus measuring the spin
correlation in $t \bar t$ events can simultaneously probe top
production and (indirectly) decay for potential effects due to new physics.
 
The measurements of the spin correlation between the top quark and the
top antiquark presented in this paper rely on angular distributions of
the top quark and top antiquark decay products. The charged leptons and the $d$-type quarks from the 
$W$ boson decays are the most sensitive spin analyzers,
and the $b$-quark from top quark decay contains some 
information about the top quark polarization, too.
Observables in the laboratory frame and in different top quark
spin quantization bases are explored. These variables are used to measure 
the coefficient $f_{\rm SM}$, which is related to the number 
       of events where the $t$ and $\bar{t}$ spins are correlated as
       predicted by the SM, assuming that \ttbar\ production consists
       of events with spin correlation as in the SM or without spin
       correlation. The measured value of $f_{\rm SM}$ is translated into the spin correlation
strength $A$, which is a measure for the number of events 
where the top quark and top antiquark spins are parallel minus the number
of events where they are anti-parallel with respect to 
a spin quantization axis, divided by the total number
of events.

The spin correlation in $t\bar{t}$ events has been studied previously
at the Tevatron and the LHC. The CDF and D0 collaborations have
performed a measurement of $A$ by exploring the angular correlations
of the charged leptons~\cite{d0dilepspin,cdfljetsspin}.  
The D0 collaboration has exploited a 
matrix element based approach~\cite{melnikovschulze} and reported
the first evidence for non-vanishing $t\bar{t}$ spin
correlation~\cite{d0dilepmespin,d0ljetsdilepmespin}. These
measurements are limited by statistical uncertainties and 
are in good agreement with the SM prediction.
Using the difference in azimuthal
angles of the two leptons from the decays of the $W$ bosons emerging from top
quarks in the laboratory frame, $\Delta \phi$, the ATLAS collaboration
reported the first observation of non-vanishing $t\bar{t}$ spin
correlation using 2.1~fb$^{-1}$ of LHC data, taken at 7 \tev\  
collision energy~\cite{atlasspin}.  
The CMS collaboration also measured spin correlations in dileptonic final
states at 7 \tev\ using angular correlations of the two charged leptons and the
$\Delta \phi$ observable with 5.0~fb$^{-1}$ of data~\cite{cmsspin},
showing good agreement with the SM prediction. 

In this paper,  measurements of \ttbar\ spin correlation using
the full 7~\tev\ data sample of $4.6$~fb$^{-1}$ collected by the ATLAS
collaboration are presented. 
Using events with one or two isolated
leptons in the final state,  spin correlations are measured using $\Delta
\phi$ between the lepton and one of the
final-state jets or between the two leptons, respectively.
Additional measurements are performed in the final state with two 
leptons, using observables that are sensitive to different types of sources
of new physics in \ttbar\ production. In particular, angular correlations
between the charged leptons from top quark decays  
in two different spin quantization bases and a ratio of
matrix elements in the dileptonic channel are also measured.

\section{The ATLAS Detector}
\label{sec:atlas}
The ATLAS experiment~\cite{ATLAS} is a multi-purpose particle physics 
detector. Its
cylindrical geometry provides a solid angle coverage close to
$4\pi$.\footnote{ATLAS uses a right-handed coordinate system, with its origin at
the nominal interaction point in the center of the detector. The
$z$-axis points along the beam direction, the $x$-axis from the 
interaction point to the center of the LHC ring, and the $y$-axis
upwards. In the transverse plane, cylindrical coordinates ($r,\phi$) are used,
where $\phi$ is the azimuthal angle around the beam
direction. The pseudorapidity $\eta$ is defined via the polar angle $\theta$
as $\eta= - \ln \tan (\theta/2)$.} 

Closest to the interaction point is the
inner detector, which covers a pseudorapidity range $|\eta|<2.5$. It
consists of multiple layers of silicon pixel and microstrip detectors
and a straw-tube transition radiation tracker (TRT).  Around the inner
detector, a superconducting solenoid provides a
2~T axial magnetic field. The solenoid is surrounded by
high-granularity lead/liquid-argon electromagnetic (EM) calorimeters and a
steel/scintillator-tile hadronic calorimeter in the central region. In the
forward region, 
endcap liquid-argon calorimeters have either copper or tungsten absorbers. 

The muon spectrometer is the outermost part of the detector. It
consists of several layers of trigger and tracking chambers organized 
in three stations. A
toroidal magnet system produces an azimuthal magnetic field to enable
an independent measurement of the muon track momenta. 

A three-level trigger system~\cite{trigger} is used for the ATLAS
experiment. The first level is purely hardware-based and 
is followed by two software-based trigger levels.

\section{Object Reconstruction}
\label{sec:objects}
In the SM, a top quark predominantly decays into a $W$ boson and a $b$-quark.
For this analysis \ttbar\ candidate events in two final states are
selected. In the dilepton final state, both $W$ bosons emerging from
top and antitop quarks decay leptonically into $e\nu_e$,
$\mu\nu_{\mu}$ or $\tau\nu_{\tau}$,~\footnote{We use the notation
  $e\nu_e$ for both $e^{+}\nu_e$ and $e^{-}\bar{\nu}_e$. The same
applies to $\mu\nu_{\mu}$ and $\tau\nu_{\tau}$.} with 
the $\tau$ lepton decaying into an electron or a muon and the respective 
neutrinos. In the single-lepton channel, one $W$~boson from the top or
antitop quark decays leptonically, while the other $W$ boson decays
into a quark-antiquark pair. 

Events are required to satisfy a  
single-electron or single-muon trigger with a minimum lepton transverse momentum
($\pt$) requirement that varies with the lepton flavor and 
the data-taking period to cope with the increasing instantaneous
luminosity. 
During the 2011 data-taking period the average number of simultaneous $pp$ 
interactions per beam crossing (pile-up) at the beginning of a fill of 
the LHC increased from 6 to 17. 
The primary hard-scatter event vertex is  
defined as the reconstructed vertex with at least five associated
tracks and  the highest sum of the squared \pt\ values of 
the associated tracks with \pt$>0.4$~\gev. 

Electron candidates~\cite{ElectronPerformance} are reconstructed from
energy deposits (clusters) in the electromagnetic calorimeter that are
associated with reconstructed tracks in the inner detector. They are
required to have a transverse energy, $\et$, greater than $25\gev$
and $|\eta_{\rm cluster}| < 2.47$,  
excluding the transition region $1.37 <|\eta_{\rm cluster}| < 1.52$ 
between sections of the electromagnetic calorimeters. The electron 
identification relies on a cut-based selection using 
calorimeter, tracking and combined variables such as those describing 
shower shapes in the EM calorimeter's middle layer,  
track quality requirements and track-cluster matching, 
particle identification using the TRT, and discrimination against 
photon conversions via a hit requirement in the inner pixel detector 
layer and information about reconstructed conversion vertices.
In addition, to reduce the background
from non-prompt electrons, i.e.~from decays of hadrons (including
heavy flavor) produced in jets, electron candidates are required
to be isolated from other activity in the calorimeter and in the
tracking system. An $\eta$-dependent 90\% efficient 
cut based on the transverse energy sum of cells around the direction of each
candidate is made for a cone of size  
$\Delta R = \sqrt{(\Delta\phi)^2 + (\Delta\eta)^2} = 0.2$, after 
excluding cells associated with the electron cluster itself.   
A further 90\% efficient 
isolation cut is made on the sum of track $\pt$ in a cone of radius
$\Delta R = 0.3$ around the electron track. 
The longitudinal impact parameter of the electron track with respect
to the event primary vertex, $z_{0}$, is required to be less
than 2 mm. 

Muon candidates are reconstructed from track segments in various
layers of the muon spectrometer and are matched with tracks found in the
inner detector. The final muon candidates are refitted using the complete
track information from both detector systems, and are required to have  
$\pt > 20\gev$ and $|\eta|<2.5$. 
Each muon candidate is required to be isolated from jets by a distance 
$\Delta R > 0.4$.
In addition, muon isolation requires that the transverse energy in the calorimeter  
within a cone of $\Delta R = 0.2$ is below $4\gev$ after excluding
the muon energy deposits in the calorimeter. Furthermore, muon isolation requires 
that the scalar sum of the track transverse momenta in a cone 
of $\Delta R = 0.3$ around the
muon candidate is less than $2.5\gev$ excluding the muon track. The efficiency of 
the muon isolation requirements depends weakly on the amount of pile-up and 
is typically 85\%.    

Jets are reconstructed from
clusters~\cite{ATLAS,JEScalibration} built from energy
deposits in the calorimeters using the anti-$k_t$
algorithm~\cite{ref:Cacciari2008,ref:Cacciari2006,ref:fastjet} with a
radius parameter $R=0.4$. 
The jets are calibrated using energy- and $\eta$-dependent calibration factors,
derived from simulations, to the mean energy of stable particles inside 
the jets. Additional corrections to account for the difference between
simulation and data are derived from in~situ techniques~\cite{JEScalibration}.

Calibrated jets with $\pt > 25\gev$ and $|\eta| < 2.5$ are selected. 
To reduce the background from other $pp$ interactions within the same
bunch crossing, the scalar sum of the $\pt$ of tracks matched to the jet 
and originating from the primary vertex must be at least 75\% of the scalar 
sum of the $\pt$ of all tracks matched to the jet.

If there are jets within a cone of $\Delta R =
0.2$ around a selected electron, the jet closest to the electron is 
discarded. This avoids double counting of electrons as jets. Finally,
electrons are removed if they are within $\Delta R = 0.4$ of a
selected jet.

Jets originating from or containing $b$-quarks are selected in
the single-lepton final state, making use of the long lifetime of $b$-hadrons. 
Variables using the properties of the secondary vertex 
and displaced 
tracks associated with the jet are combined by a neural network used
for $b$-jet identification~\cite{btagbasic}.  
A working point with a 70\% $b$-tagging efficiency is used to select \ttbar\
events~\cite{btagging} in the single-lepton channel. 

The magnitude of the missing transverse momentum ($\met$) is reconstructed from the 
vector sum of all calorimeter cell energies associated with topological 
clusters with $|\eta| < 4.5$~\cite{METcalibration}. Contributions from
the calorimeter energy clusters matched with either a reconstructed
lepton or jet are corrected to the corresponding energy scale. The
term accounting for the \pt\ of any selected muon is included in the
$\met$ calculation.

\section{Event Selection}
\label{sec:selection}
\subsection{Dilepton channel}
To select \ttbar\ candidate events with leptonic $W$ decays, two
  leptons  
of opposite charge ($\ell^+\ell^-$ = \ee, \mumu\ or \emu) and at least two
 jets are required. 
For the \mumu\ final state, events containing a muon pair consistent with a 
cosmic-ray muon signature are rejected. In particular,   
events are rejected if two muon tracks are back to back in $\phi$, 
they have the same sign pseudorapidity, and the
point of closest approach to the primary vertex of each track is
greater than 5 mm.
Since the same-flavor leptonic channels \ee\ and \mumu\  
suffer from a large 
background from the leptonic decays of hadronic resonances, such as the
$J/\psi$ and $\Upsilon$, the invariant mass of the
two leptons, \mll, is required to be larger than 15~\gev. 
A contribution from the Drell--Yan production of $Z/\gamma^*$ bosons 
in association with jets ($Z/\gamma^*$+jets production) to these channels 
is suppressed by 
rejecting events where \mll\ is close to the $Z$ boson mass $m_Z$; i.e. $|\mll - m_Z| >
10$~\gev\ is required. In addition, large missing transverse momentum, \met\ $>$
60 \GeV, is required to account for the two neutrinos from the leptonic 
decays of the two $W$~bosons. 
Events with at least two jets, $|\mll - m_Z| < 10$~\gev, and 
\met\ $>$ 30 \GeV\  are used as a control region to validate modeling
of the spin observables (see Sec.~\ref{sec:measurement_dil}).  
 
The \emu\ channel  does not suffer from an overwhelming Drell--Yan
background. Therefore the \mll\ cut is not applied. 
To suppress the remaining background from  $Z/\gamma^*(\to
\tau^{+}\tau^{-})$+jets production  a cut on the scalar sum of the transverse
energy of leptons and jets, \mbox{\htlep\ $>$ 130~\gev}, is applied instead of
the \met\ cut. The purity of the \ttbar\ sample after the dilepton selection 
is about 85\%.    

\subsection{Single-lepton channel}
To select \ttbar\ candidate events in the single-lepton final state, exactly
one isolated lepton (electron or muon), at least four jets and high
\met\ are required. 
The \met\ has to be larger than 30~GeV (20~GeV) in the 
$e$+jets ($\mu$+jets) final state to account for the
neutrino from the leptonic decay of a $W$ boson. 
To suppress the contribution of QCD multijet events a cut on the
$W$~boson transverse mass,~\footnote{In events with a leptonic decay of a $W$ 
boson, $m_{\textrm{T}}(W) = \sqrt{2 \pt^{\ell}
    \pt^\nu ( 1 - \cos (\phi^{\ell} - \phi^\nu))}$ where $\pt^{\ell}$ and 
  $\pt^\nu$ ($\phi^{\ell}$ and $\phi^\nu$) are the transverse momenta
  (azimuthal angle) of the charged lepton and neutrino,
  respectively. The measured $\met$ vector provides the neutrino information.}
$m_{\textrm{T}}(W)>30$~GeV, is applied in the $e$+jets final state while 
in the $\mu$+jets final state, 
$m_{\textrm{T}}(W)+\met$ is required to be larger than 60~GeV. 
In both channels, at least one of the jets has to be identified as a 
$b$-jet by the $b$-tagging algorithm,  resulting in a 78\% ($e$+jets)
and 76\% ($\mu$+jets) pure \ttbar\ sample.

\section{Sample Composition and Modeling}
\label{sec:sample}
After event selection, the sample is composed of $t\bar{t}$ signal and
various backgrounds. In the following, the sample composition of the
dilepton and single-lepton channels are discussed. 

\subsection{Dilepton channel} \label{sec:dilepsample}
\begin{table*}[htbp]
  \caption{Observed numbers of events in data compared to the
    expectation
       after the selection in the dilepton channels.
     Backgrounds and signal estimated from simulation are indicated
     with the (MC) suffix,
     whereas backgrounds estimated using data-driven techniques are
     indicated with
     a (DD) suffix. Quoted uncertainties include the
     statistical uncertainty on the yield and the uncertainty
     on the theoretical cross sections used for MC normalization. The
     uncertainty on
     the DD estimate is statistical only.}
\label{tab:dilep_cutflow}
\begin{ruledtabular}
\begin{tabular}{cccc}
& \ee & \mumu & \emu \\ \hline
$Z(\rightarrow \ell^{+} \ell^{-})+$jets (DD/MC) & 21 \ppm\ 3 & \hphantom{0}83 \ppm\ 9\hphantom{0}  & --- \\
$Z(\rightarrow \tau^{+} \tau^{-})$+jets (MC) & 18 \ppm\ 6 & \hphantom{0}67 \ppm\ 23             & 172 \ppm\ 59 \\
Fake leptons (DD)                           & 20 \ppm\ 7 & \hphantom{0}29 \ppm\ 4\hphantom{0} & 101 \ppm\ 15 \\
Single top (MC)                             & 31 \ppm\ 3 & \hphantom{0}83 \ppm\ 7\hphantom{0} & 224 \ppm\ 17 \\
Diboson (MC)                                & 23 \ppm\ 8 & \hphantom{0}60 \ppm\ 21            & 174 \ppm\ 59 \\
\hline
Total (non-\ttbar) & 112 \ppm\ 13 & 322 \ppm\ 33 & 671 \ppm\ 87 \\
\ttbar\ (MC) & 610 \ppm\ 37 & 1750 \ppm\ 110 & 4610 \ppm\ 280 \\
\hline
Expected & 721 \ppm\ 39 & 2070 \ppm\ 110 & 5280 \ppm\ 290 \\
Observed & 736 & 2057 & 5320 \\
\end{tabular}
\end{ruledtabular}
\end{table*}

Backgrounds to same-flavor dilepton \ttbar\ production arise 
from the Drell--Yan $Z/\gamma^*$+jets production process with 
the $Z/\gamma^*$ boson decaying into 
$e^{+}e^{-}$ or  $\mu^{+}\mu^{-}$. In the \emu\ channel, one of the main
backgrounds is due to $Z/\gamma^*$+jets production with decays
$Z/\gamma^* \to \tau^{+}\tau^{-}$,
followed by leptonic decays of the $\tau$ leptons. 
Other backgrounds in dilepton channels are due to diboson  
production, associated production of a single top quark and a
$W$~boson ($Wt$), \ttbar\ production with a 
single-lepton in the final state, single top quark production via $s$-
or $t$-channel exchange of a $W$ boson, and the production of a $W$ boson 
in association with jets.   
The latter three processes contain non-prompt leptons that pass the lepton isolation requirement or misidentified leptons arising from jets.   
The contributions from these
processes are estimated using data-driven methods.

Drell--Yan events are generated using 
the {\sc Alpgen v2.13}~\cite{alpgen} generator including leading-order
(LO) matrix
elements with up to five additional   
partons. The {\sc CTEQ6L1} parton distribution function (PDF)
set~\cite{cteq6} is used, and the cross 
section is normalized to the next-to-next-to-leading-order (NNLO)
prediction~\cite{ZWPROD}. 
Parton showering and
fragmentation are modeled by {\sc Herwig} v6.520~\cite{herwig}, and the
underlying event is simulated by {\sc Jimmy}~\cite{jimmy}. To avoid
double counting of partonic configurations generated by both the
matrix-element calculation and the parton-shower evolution, a
parton--jet matching scheme (``MLM matching")~\cite{mlm} is employed.  
The yields of dielectron and dimuon Drell--Yan events predicted by the 
Monte Carlo (MC) simulation are compared to the data in 
$Z/\gamma^*$+jets-dominated control
regions. Correction factors are derived and applied to the predicted
yields in the signal region, to account for the difference between 
the simulation 
prediction and data. The correction increases the $Z/\gamma^*
\rightarrow e^{+}e^{-}$+jets contribution by 3\%  
and the  $Z/\gamma^*
\rightarrow \mu^{+}\mu^{-}$+jets contribution by 13\% 
relative to the prediction from simulation. 

Single top quark background arises from the associated $Wt$ production,  
when both the $W$ boson
emerging from the top quark and the $W$ boson from the hard
interaction decay leptonically. This contribution  is 
modeled with \mcatnlo\ v4.01~\cite{mcatnlo_1,mcatnlo_2,mcatnlo_3} 
using the {\sc
CT10} PDF set~\cite{ct10} and normalized to the approximate 
NNLO theoretical cross section~\cite{stop}.   
 
Finally, the diboson backgrounds are modeled using {\sc Alpgen v2.13}
interfaced with {\sc Herwig} using the {\sc MRST LO**} PDF
set~\cite{mrst} and   
normalized to the theoretical calculation at next-to-leading-order
(NLO) in QCD~\cite{dibosonxs}. 

The background arising from the misidentified and non-prompt leptons
(collectively referred to as ``fake leptons'') is determined from
data using the ``matrix method'', which was
previously used in the measurement described in Refs.~\cite{matrixmethod1,
  matrixmethod2}. 

The SM \ttbar\ signal events are modeled using the \mcatnlo\ v4.01
generator. Top quarks and the subsequent $W$ bosons are decayed
conserving the spin correlation information. The decay products are interfaced 
with {\sc Herwig}, which showers the $b$ quarks and $W$ boson daughters,
and with {\sc Jimmy} to simulate multiparton interactions.
A top quark mass of $172.5\gev$ is assumed.
The {\sc CT10} PDF set is used. 

The generation chain can be 
modified such that top quarks are decayed by {\sc Herwig} rather than \mcatnlo.
In this case the top quark spin information is not propagated to the
decay products, and therefore the spins between the top quarks
are uncorrelated.    
This technique has a side effect that the top quarks in the uncorrelated case are  
treated as being on-shell, and hence they do not have an intrinsic width. The effect 
of this limitation is found to be negligible. 

All MC samples use a {\sc Geant4}-based simulation to model the ATLAS
detector~\cite{geant4,atlassim}. For each MC process,  
pile-up is overlaid  using
simulated  
minimum-bias events from the {\sc Pythia} generator. The number of 
additional $pp$ interactions 
is reweighted to the number of interactions observed in data.

In Table~\ref{tab:dilep_cutflow} the observed yields in data are compared
to the expected background and the \ttbar\ signal normalized to 
$\sigma_{t\bar{t}}= 177^{+10}_{-11}$~pb calculated at NNLO  
in QCD including resummation of next-to-next-to-leading logarithmic 
soft gluon terms~\cite{xs1,xs2,xs3,xs4,xs5} with {\sc Top++}
v2.0~\cite{xs6} for a top quark mass of $172.5 \gev$. 
A significantly lower yield in the dielectron channel compared to the dimuon channel is
due to the stringent isolation criteria and higher \pt\ cut on the electrons.  
The yield difference between \ttbar\ signal with SM spin correlation and without 
spin correlation is found to be negligible in the $e^{\pm}\mu^{\mp}$
channel but not in the $e^+e^-$ or $\mu^+\mu^-$ channels. Here, the cut on
the invariant mass of the dilepton system used to suppress
backgrounds also preferentially selects uncorrelated
\ttbar\ pairs over correlated pairs. This is due to the fact that 
on average uncorrelated \ttbar\ pairs have larger values of 
$\Delta\phi (\ell,\ell)$, which translates into larger values of 
\mll\ and therefore more events passing the $|\mll - m_Z| > 10$~\gev\ selection cut.   
This effect is accounted for in the analysis.    

\subsection{Single-lepton channel}
In the single-lepton channel the main background is due to $W$+jets
production, where the $W$~boson 
decays leptonically. Other background contributions arise from 
$Z/\gamma^*$+jets production, where the $Z$ boson decays
into a pair of leptons and one of the leptons does not pass the selection 
requirements, from electroweak processes (diboson
and single top quark production in the $s$-, $t$- channel, and $Wt$-processes) and 
from multijets events, where a lepton from the decay of a heavy-flavor quark
appears isolated or a jet mimics an electron.  
Additional background arising from $t\bar{t}$ events with two
leptons in the final state, where one lepton lies outside the
acceptance, is studied with \mcatnlo\ MC simulation and treated as part of the 
signal. 
The diboson, single top quark and $Z/\gamma^*$+jets backgrounds are
estimated using simulated events  normalized to the theoretical cross 
sections. The $W$+jets events are 
generated with {\sc Alpgen v2.13}, using the {\sc CTEQ6L1} PDF
set with up to five additional partons. Separate samples are
generated for $W+b\bar{b}$, $W+c\bar{c}$ and $W+c$ production at the 
matrix-element level. The normalization of the $W$+jets background and its 
heavy-flavor content are extracted from 
data by a method exploiting the $W$+jets production charge 
asymmetry~\cite{matrixmethod1}. 
Single top quark $s$-channel  and $Wt$-channel production is
generated using \mcatnlo, where the  diagram removal scheme is invoked
in the $Wt$-channel production to avoid 
overlap between single top quark and $t\bar{t}$ final states~\cite{Frixione:2008yi}. 
For the $t$-channel, 
{\sc AcerMC}~\cite{acermc} with {\sc Pythia} parton shower and modified LO 
PDFs (\textsc{MRST LO**}~\cite{Sherstnev:2007nd}) is used.

The QCD multijet background is estimated from data using the
same matrix method as in the dilepton 
channel~\cite{matrixmethod1,matrixmethod2}.

Table~\ref{tab:ljets_cutflow} shows the observed yields in data, compared
to the expectation from the background and the \ttbar\ signal. The
expectation is in good agreement with the data. 

\begin{table}[htbp]
   \caption{Observed numbers of events in data compared to the
      expectation
       after the selection in the single-lepton channels.
     Backgrounds and signal are estimated from simulation (MC) or from
     data-driven techniques (DD). Quoted uncertainties include the
     statistical uncertainty on the yield and the uncertainty
     on the theoretical cross sections used for MC normalization. The
     uncertainty on
     the DD estimate is statistical only.}
\begin{ruledtabular}
\begin{tabular}{ccc}
$n_{\text{jets}} \geq 4$, $n_{\text{$b$-tags}} \geq 1$ & \ejets\ & \mujets\ \\
\hline
$W+$jets (DD/MC)   &         \hphantom{0}2320 \ppm\  390 &  \hphantom{0}4840 \ppm\  770\\
$Z+$jets (MC)      &         \hphantom{00}450 \ppm\  210 &  \hphantom{00}480 \ppm\  230\\
Fake leptons (DD)  &         \hphantom{00}840 \ppm\  420 &  \hphantom{0}1830 \ppm\  340\\
Single top (MC)    &                     1186 \ppm\   55 &              1975 \ppm\   83\\
Diboson (MC)       &           \hphantom{0}46 \ppm\    2 &    \hphantom{0}73 \ppm\    4\\
\hline
Total (non-\ttbar) &         \hphantom{0}4830 \ppm\  620 &  \hphantom{0}9200 \ppm\  890\\
\hline
\ttbar\ (MC, $\ell$+jets) &             15130 \ppm\  900 & \hphantom{0}25200 \ppm\ 1500\\
\ttbar\ (MC, dilepton) &     \hphantom{0}2090 \ppm\  120 &  \hphantom{0}3130 \ppm\  190\\
\hline
Expected &                  \hphantom{0}22100 \ppm\ 1100 & \hphantom{0}37500 \ppm\ 1800\\
Observed & \hphantom{0}21770 & \hphantom{0}37645\\
\end{tabular}
\end{ruledtabular}
\label{tab:ljets_cutflow}
\end{table}

\section{Spin Correlation Observables}
\label{sec:observables}
The spin correlation of pair-produced top quarks is extracted
from the angular distributions of the top quark decay products
in $t \to Wb$ followed by $W \to \ell \nu$ or $W \to q \bar{q}$. The
single differential angular 
distribution of the top decay width $\Gamma$ is given by
\begin{equation}\label{eq:topdecayl}
 \frac{1}{\Gamma} \frac{d\Gamma}{d\cos(\theta_i)} = (1+ \alpha_i {\bf |P|}
 \cos(\theta_i))/2 \, ,
\end{equation}
where $\theta_i$ is the angle between the momentum direction of 
decay product $i$ of the top (antitop) quark
and the top (antitop) quark polarization three-vector $\bf P$, $0 \le
|{\bf P}| \le 1$. The factor  $\alpha_i$ is the spin-analyzing 
power, which must be between $-1$ and $1$. 
At NLO, the factor  $\alpha_i$  is
predicted to be $\alpha_{\ell^+}=+0.998$ for positively charged
leptons~\cite{polpower,Bernreuther:2001rq,Parke_Top2011},
$\alpha_{d} = -0.966$ for down
quarks, 
$\alpha_b = -0.393$ for bottom
quarks~\cite{polpower,Brandenburg:2002xr,Parke_Top2011}, and
the same $\alpha_{i}$ value with opposite sign for the respective antiparticles.

In the SM, the 
polarization of the pair-produced top quarks in $pp$ collisions 
is negligible~\cite{bernreuther_si_chromo}. Ignoring it, 
the correlation between the decay products of the top quark (denoted with 
subscript $+$) and the top antiquark (denoted with 
subscript $-$) can be expressed by
\begin{equation}
 \frac{1}{\sigma} \frac{d\sigma}{d\cos(\theta_{+}) \ d\cos(\theta_-)}=
 \frac{1}{4}\left( 1 + A \, \alpha_+ \alpha_- \cos(\theta_+)  \cos(\theta_-)  \right) \, , 
\label{eq:coscos}
\end{equation}
with  
\begin{equation}
    A = \frac{{N_{\mathrm {like}}} - N_{{\mathrm {unlike}}}}{{N_{\mathrm {like}}} + 
    N_{\mathrm {unlike}}} =
     \frac{N(\uparrow \uparrow) + N(\downarrow \downarrow) - N(\uparrow \downarrow) - N(\downarrow \uparrow)}
     {N(\uparrow \uparrow) + N(\downarrow \downarrow) + N(\uparrow \downarrow) +
     N(\downarrow \uparrow)} \, ,
\label{eq:A}
\end{equation}
where ${N_{\mathrm {like}}}=N(\uparrow \uparrow)+N(\downarrow \downarrow)$ is 
the number of events where the top quark and top antiquark spins are
parallel, and $N_{{\mathrm {unlike}}}=N(\uparrow \downarrow)+N(\downarrow \uparrow)$
is the number of events where they are anti-parallel with respect to the spin quantization axis. The strength 
of the spin correlation is defined by 
\begin{equation}\label{eq:C}
    C = - A \, \alpha_{+} \alpha_{-} \,.
\end{equation}
Using the mean of the doubly differential cross section in 
Eq.~(\ref{eq:coscos}), $C$ can be extracted as
\begin{equation}\label{eq:C_expect}
C = -9 \, \langle  \cos(\theta_+)  \cos(\theta_-) \rangle .
\end{equation}
In this paper, however, the full distribution of 
$\cos(\theta_+)  \cos(\theta_-)$, as defined in
Eq.~(\ref{eq:coscos}), is used.  In dilepton final
states where the spin-analyzing power is effectively 100\%, $C
\approx A$. 
To allow for a comparison to previous analyses, the
results are given both in terms of $f_{\rm SM}$ defined in Sec.~\ref{sec:extractspin}, 
and in terms of $A$.  

Four observables are used to extract the spin correlation
strength. The first variable is used in both the dilepton and the single-lepton
final states, and the latter three variables are only used in the dilepton final state.
\begin{itemize}
\item The azimuthal opening angle,  $\Delta\phi$, between the momentum
  directions of a top quark decay
  product and an anti-top quark decay product in the laboratory
  frame. In the dilepton final state, $\Delta\phi$ between  
  the charged lepton momentum directions, 
  $\Delta\phi(\ell^{+},\ell^{-})$, is explored. This observable is 
  straightforward to measure and very sensitive because like-helicity 
  gluon--gluon initial states dominate~\cite{parke_deltaphi}.   
  It was used in
  Ref.~\cite{atlasspin} to observe a non-vanishing spin correlation,
  consistent with the SM prediction.
In the single-lepton channel, $\Delta\phi$ between the charged lepton
momentum direction and either the down-type jet from $W$ boson decay, 
$\Delta\phi(\ell,d)$, or the $b$-jet from the hadronically decaying
top quark, $\Delta\phi(\ell,b)$, are
analyzed. Since this requires the identification of the jets from the 
$W$ boson and hadronically decaying top quark, full event
reconstruction is necessary, making the measurement of $\Delta \phi$ in 
the single-lepton channel more challenging. Moreover, 
there is a need to 
identify the jet emerging from the down-type quark from $W$ boson
decay (see Sec.~\ref{sec:ljetskin} for more details).  
\item The ``$S$-ratio'' of matrix elements ${\cal M}$ for top quark
  production and decay from the fusion of like-helicity gluons 
  ($g_{\mathrm R} g_{\mathrm R} + g_{\mathrm L} g_{\mathrm L}
  \rightarrow t \bar{t} \rightarrow (b  \ell^{+} \nu) (\bar{b}  \ell^{-} \bar{\nu})$ ) with SM spin
  correlation and without spin correlation at LO~\cite{parke_deltaphi},
\begin{eqnarray}
\label{eq:sratio}
  S &=& \frac{(|{\cal M}|_{\rm RR}^2 +
    |{\cal M}|_{\rm LL}^2)_{\mathrm {corr}}}{(|{\cal M}|_{\rm RR}^2 +
    |{\cal M}|_{\rm LL}^2)_{\mathrm {uncorr}}}\\ \nonumber
     &=& \frac{m_t^2\{(t \cdot \ell^{+})  (t \cdot \ell^{-}) + (\tbar \cdot
       \ell^{+})  (\tbar \cdot \ell^{-}) - m_t^2  (\ell^{+} \cdot
       \ell^{-})\}}{(t \cdot \ell^{+})(\tbar \cdot \ell^{-})(t \cdot \tbar)
     }\, ,
\end{eqnarray}
where $t$, $\bar{t}$, $\ell^+$, and $\ell^-$ are the 4-momenta of the top
quarks and the charged leptons. Since the like-helicity gluon--gluon 
matrix elements are used for the construction of  the $S$-ratio, it is 
particularly sensitive to like-helicity gluon--gluon initial states.
To measure this observable, and the two others described below, the
top quark and the top antiquark have to be fully reconstructed.
\item The double differential distribution (Eq.~(\ref{eq:coscos})),
  where the top direction in the $t\bar{t}$ rest 
  frame is used as the spin quantization axis.
The measurement of this distribution allows for a direct
extraction of the spin correlation strength
$A_{\rm helicity}$~\cite{bernreutherpol}, as defined in Eq.~(\ref{eq:A}).
The SM prediction is
$A_{\rm helicity}^{\rm SM}=0.31$, which was calculated including NLO QCD
corrections to \ttbar\ production and decay and mixed weak-QCD
corrections to the production amplitudes in
Ref.~\cite{bernreuther_si_chromo}. 
\mcatnlo, which includes NLO QCD corrections to \ttbar\ production 
but not to top quark decay, reproduces the same value after  
adding parton shower simulated by {\textsc{Herwig}}. 
\item The double differential distribution (Eq.~(\ref{eq:coscos})), using 
  the ``maximal'' basis as the top spin quantization axis. For the gluon--gluon fusion
  process, which is a mixture of like-helicity and unlike-helicity
  initial states, no optimal axis exists where the spin correlation 
  strength is $100\%$. This is in contrast to the quark-antiquark
  annihilation process where an optimal ``off-diagonal'' basis was
  first identified by Ref.~\cite{opt_basis}.
   However, event by event a quantization axis that maximizes spin 
   correlation and is called the ``maximal'' basis can be constructed for the
  gluon fusion process~\cite{maxbasis}. 
A prediction for the \ttbar\ spin correlation using this observable is not 
yet available for  7~\TeV $pp$ collisions. Therefore,
  the prediction is calculated using the \mcatnlo+{\textsc{Herwig}} simulation 
  resulting in $A_{\rm maximal}^{\rm SM}=0.44$.
\end{itemize}

\begin{figure*}[htpb!]
\begin{center}
\subfigure[]{\includegraphics[width=0.49\textwidth]{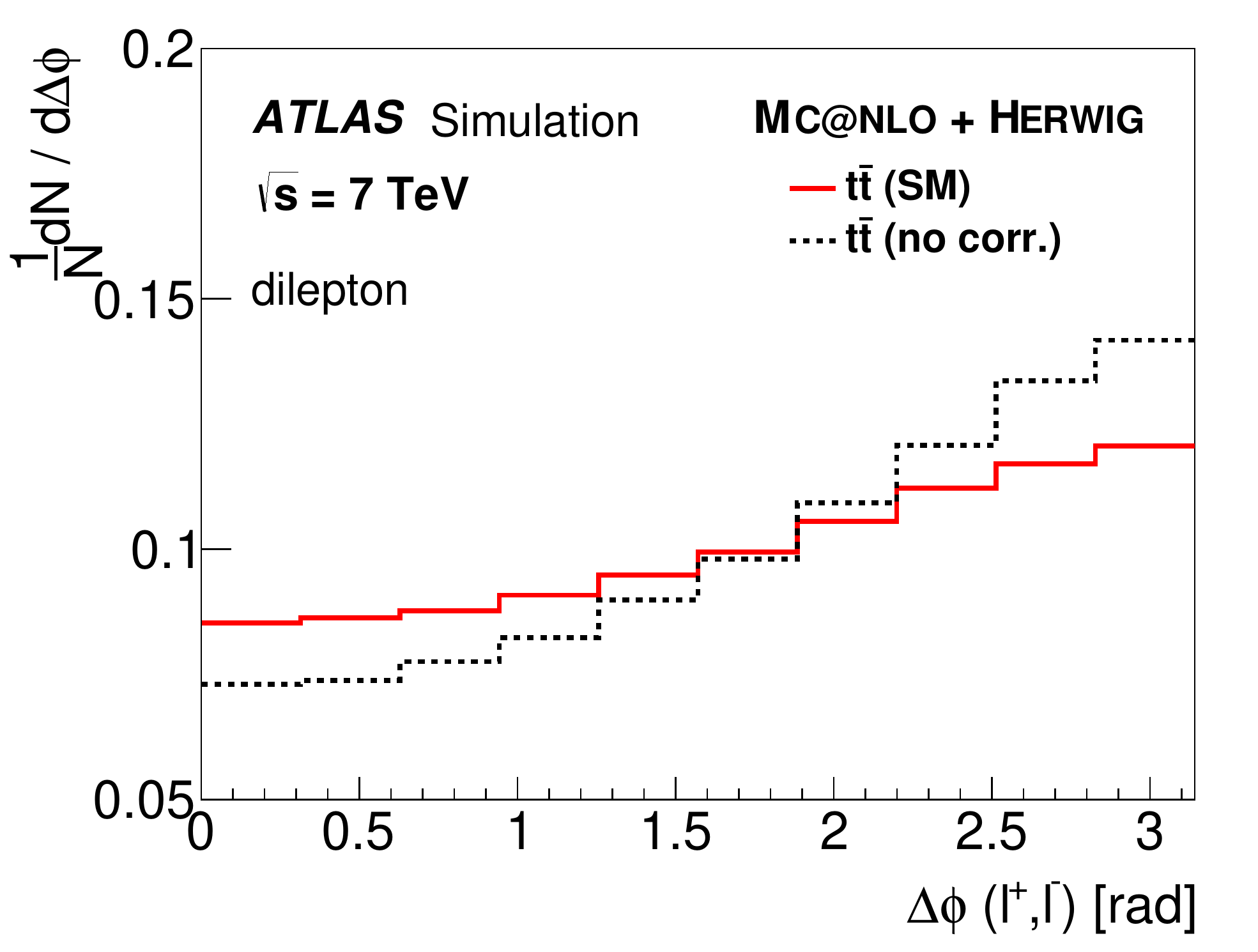}}\label{fig:obs_parton_a}
\subfigure[]{\includegraphics[width=0.49\textwidth]{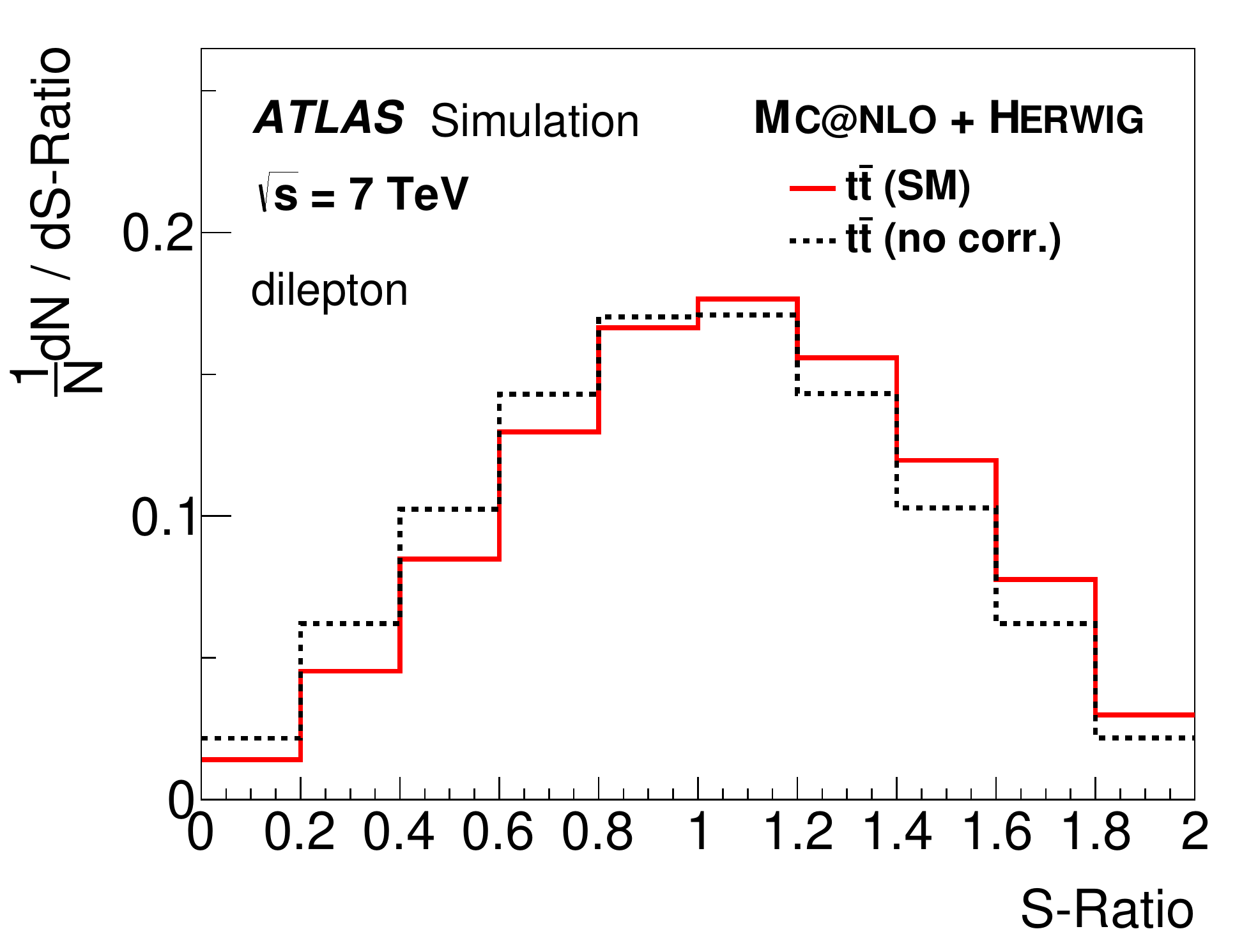}}\label{fig:obs_parton_b}\\
\subfigure[]{\includegraphics[width=0.49\textwidth]{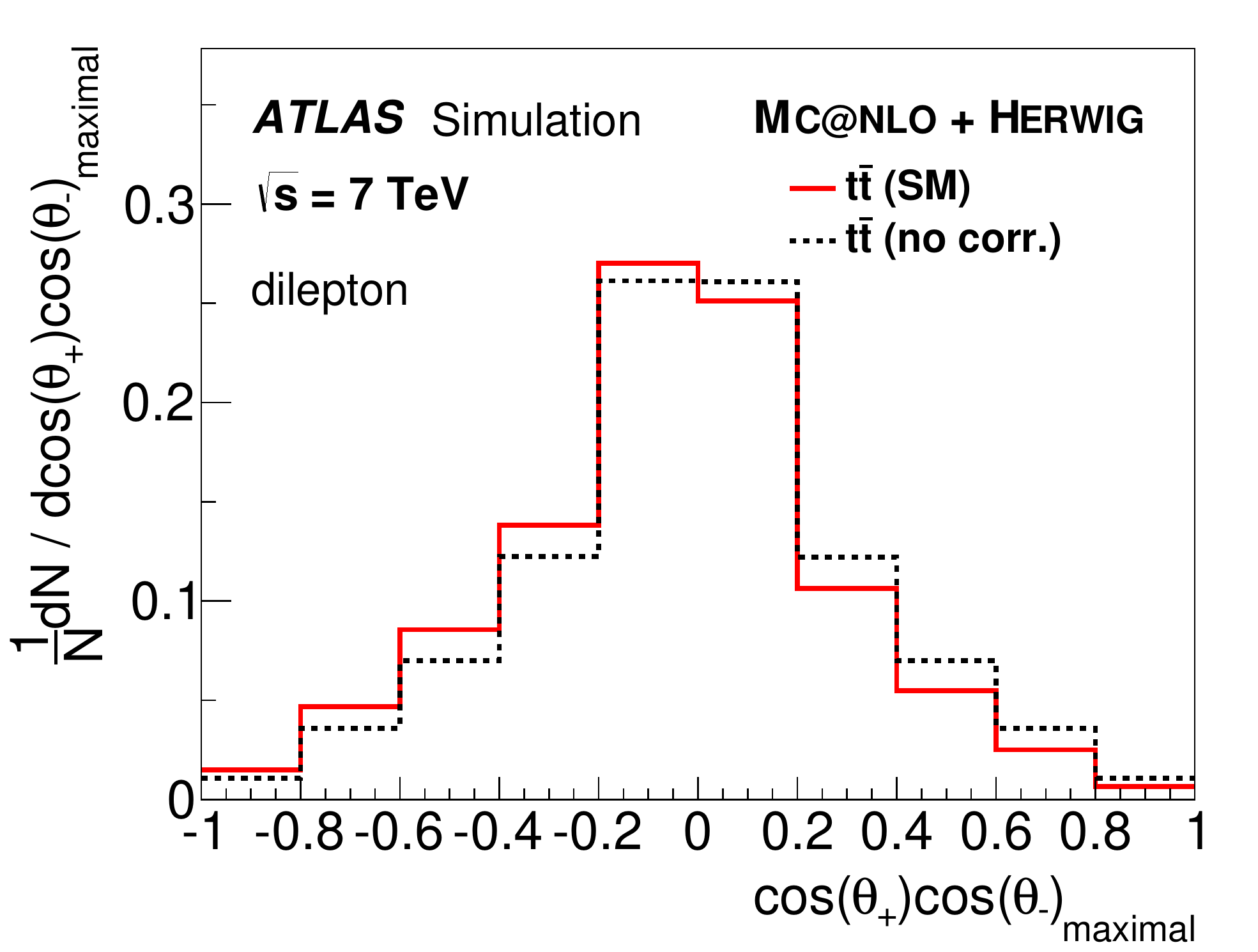}}\label{fig:obs_parton_c}
\subfigure[]{\includegraphics[width=0.49\textwidth]{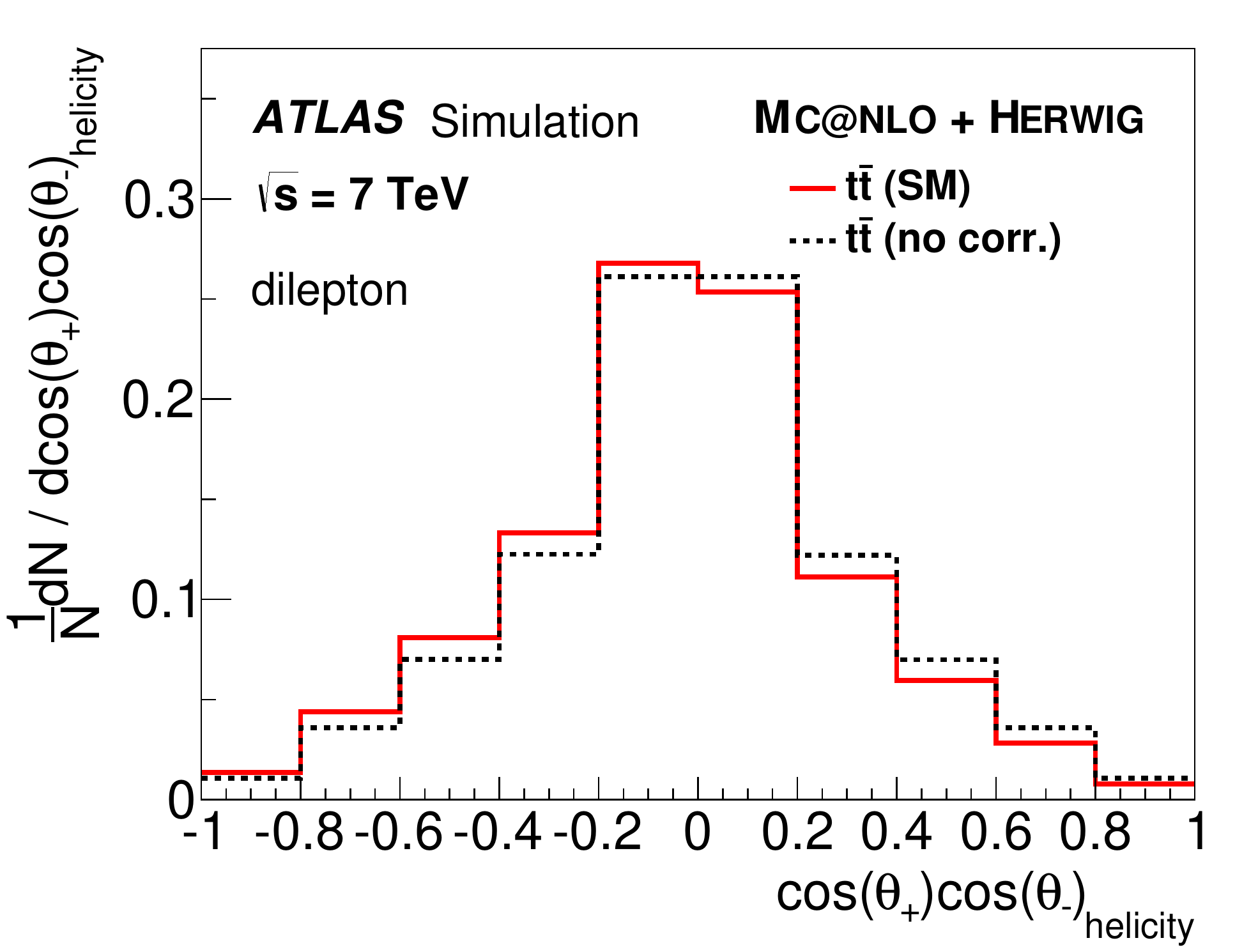}}\label{fig:obs_parton_d}
\end{center}
\caption{Distributions of several observables for generated charged
  leptons from top quark 
decay and top quarks: (a) $\Delta\phi(\ell^{+},\ell^{-})$; (b) $S$-ratio, as
    defined in Eq.~(\ref{eq:sratio}); (c)
    $\cos(\theta_{+})\cos(\theta_{-})$, as defined in
    Eq.~(\ref{eq:coscos}) in the helicity basis; (d) in the 
    maximal basis.  
    The normalized distributions show predictions for
  SM spin correlation (red solid lines) and no spin correlation (black
  dotted lines). }
\label{fig:obs_parton}
\end{figure*}

Figure~\ref{fig:obs_parton} shows all four
observables for (a) generated charged leptons from top quark decay 
and (b,c,d) top quarks in the dilepton final state, calculated with \mcatnlo\ 
under the
assumption of SM \ttbar\ spin correlation and no spin correlation, as 
defined in Sec.~\ref{sec:sample}.

The measurement of the four variables in the dilepton final state does not
comprise redundant information. It can be shown that the hadronic
$t\bar{t}$ production density matrices at tree level can be decomposed
into different terms analyzing top quark spin-independent effects, top quark
polarization, and $t\bar{t}$ spin correlations~\cite{bernreuther_matrices}. 
Using rotational invariance, these terms can be structured according to
their discrete symmetry properties. In this way four independent
$C$-even and $P$-even spin correlation coefficients that are functions of the partonic
center-of-mass energy and the production angle are introduced. The
four observables investigated here depend on different linear
combinations of these four coefficient functions.

\begin{figure*}[htpb!]
\begin{center}
\subfigure[]{\includegraphics[width=0.49\textwidth]{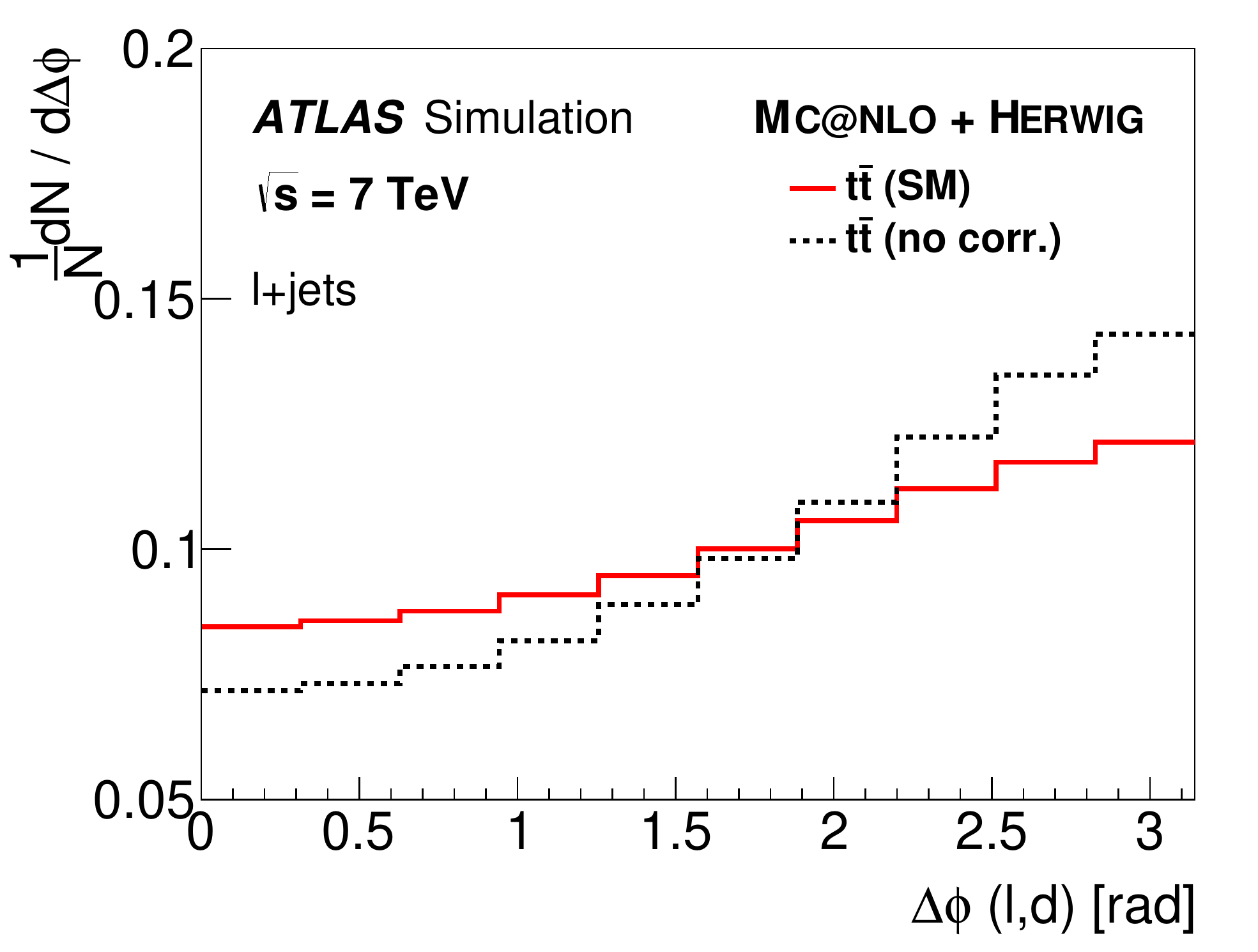}}
\subfigure[]{\includegraphics[width=0.49\textwidth]{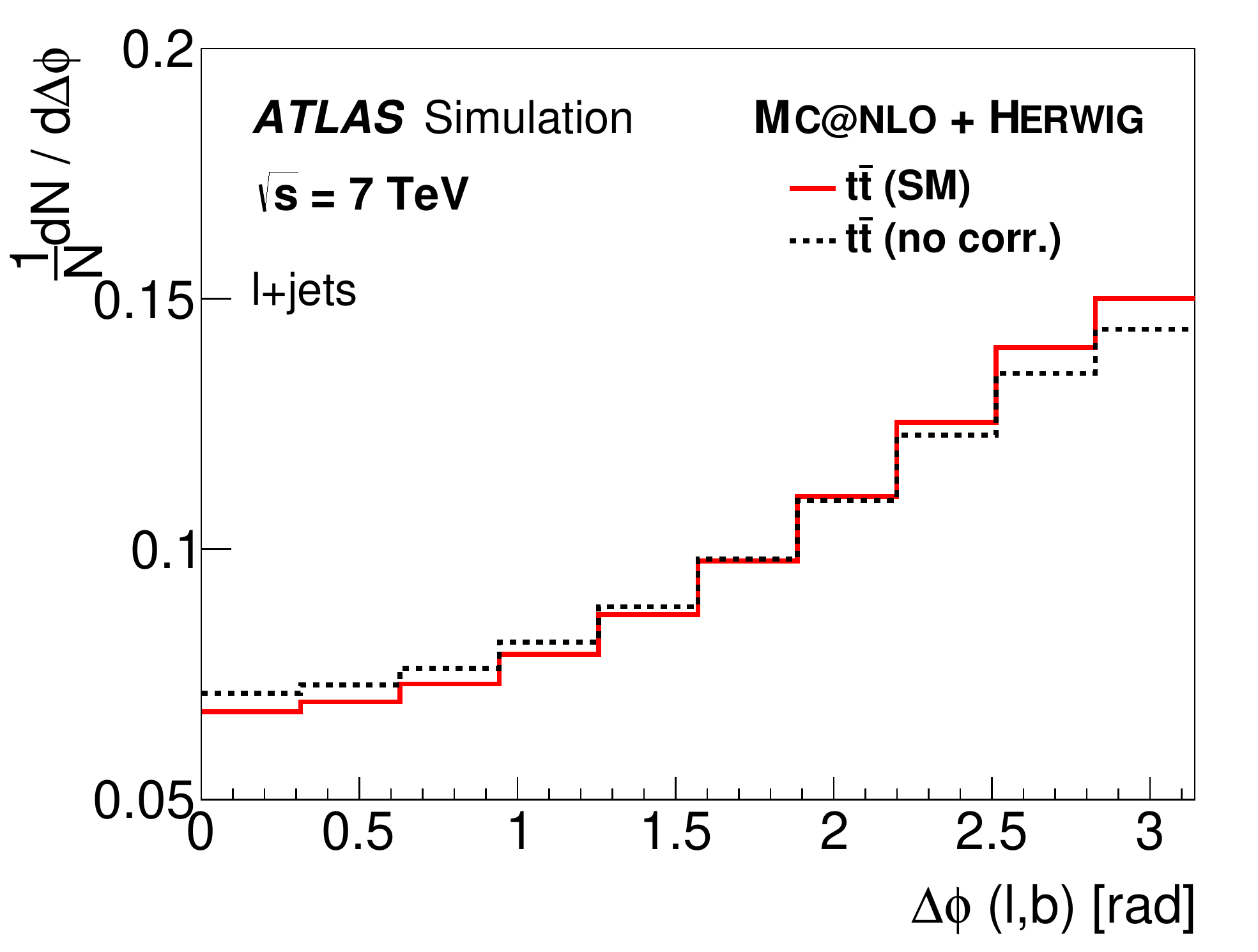}}\\
\end{center}
\caption{Distribution of $\Delta\phi$: (a) between lepton and $d$-quark; (b) 
between lepton and $b$-quark, for generated top quark decay products. The 
normalized distributions show predictions for SM spin correlation (red
solid lines) and no spin correlation (black dotted lines). }
\label{fig:obs_parton_lj}
\end{figure*}

In the single-lepton final state, $\Delta\phi(\ell,d)$ and
$\Delta\phi(\ell,b)$ are used in the analysis. Their distributions are
shown in Fig.~\ref{fig:obs_parton_lj} for generated leptons and quarks and 
are identical in the absence of spin correlation. The presence of 
spin correlation causes a split into two distributions such that 
the $\Delta\phi(\ell,b)$ distribution becomes steeper while 
the trend is opposite for $\Delta\phi(\ell,d)$.
At parton level the separation between the distribution
with SM spin correlation and without spin correlation 
for $\Delta\phi(\ell,d)$ is 
similar to the one for $\Delta\phi(\ell,\ell)$ in the dilepton channel
while the separation is significantly smaller for
$\Delta\phi(\ell,b)$.  

\section{Measurement Procedure}
 \label{sec:measurement}
After selecting a $t\bar{t}$-enriched data sample and estimating 
the signal and background composition, the spin correlation
observables, as defined in Sec.~\ref{sec:observables}, are measured and
used to extract the strength of the \ttbar\ spin correlation. 

In the dilepton final state, the $\Delta \phi(\ell,\ell)$ observable is
the absolute value of the difference in $\phi$ of the two leptons,
i.e. it is measured in the laboratory frame.
Figure~\ref{fig:ee_DY}a and~\ref{fig:mumu_DY}a show
this distribution in the \ee\ and \mumu\ channels, respectively, 
in a control region dominated by $Z/\gamma^*$+jets production. This
region is selected using the same requirements as for the signal
sample selection, but inverting the $Z$ mass window cut, defined in
Sec.~\ref{sec:selection}.
The other observables in the dilepton final state, $\cos(\theta_{+}) \cos(\theta_{-})$ and the
$S$-ratio, require the
reconstruction of the full kinematics of the \ttbar\ system discussed in 
Sec.~\ref{sec:measurement_dil}. 

In the single-lepton final state,  two observables for the spin
correlation measurement are used, \ddq\
and \dbq, that both require event reconstruction to identify the jets
from $W$-boson and top quark decay. Furthermore, a larger sensitivity
to the 
modeling of the kinematics of $t\bar{t}$ events requires a somewhat different
approach than in the dilepton final state: instead of fitting \ddq\
and \dbq\ separately, a fit to the combination of both observables is used. 

\subsection{Kinematic reconstruction of the \ttbar\ system in the
  dilepton final state}
\label{sec:measurement_dil}

The two neutrinos from $W$-boson decays in dilepton final states
cannot be measured but can only be inferred from the measured missing
transverse momentum in the event. Since only the sum of the missing
transverse momenta of the two neutrinos is measured, the system is
underconstrained. 

In this analysis a method known as 
the ``neutrino weighting technique''~\cite{NuWeighting} 
is employed. 
To solve the event kinematics
and assign the final-state objects to the decay products of the top
quark and top antiquarks, the
invariant mass calculated from the reconstructed charged  lepton and
the assumed neutrino has
to correspond to the $W$-boson mass, and the invariant mass of
the jet--lepton--neutrino combination is constrained to the top
quark mass. 
To fully solve the kinematics, the
pseudorapidities $\eta_1$ and $\eta_2$ of the two neutrinos 
are sampled from a fit of a Gaussian function to the respective distributions in
a simulated sample of $t\bar{t}$ events. It was verified that the $\eta_1$
and $\eta_2$ distributions in $t\bar{t}$ events do not change for
different \ttbar\ spin correlation strengths. 
Fifty values are chosen for each neutrino $\eta$, with $-4<\eta_{1,2}<4$
taken independently of each other. 

By scanning over all $\eta_1$ and $\eta_2$ configurations taken from the 
simulation, all possible
solutions of how to assign the charged leptons, neutrinos and
jets to their parent top quarks are accounted for. In
addition, the energies of the reconstructed jets are smeared according to the 
experimental resolution~\cite{jer}, and
the solutions are re-calculated for every smearing step. If no
solution is found, the event is discarded. 
Around 95\% of simulated \ttbar\ events have at least one
solution. This fraction is considerably lower for the backgrounds,
leading to an increase by 25\% in the signal-to-background ratio  when
requiring at least one solution.
Each solution is assigned a weight, defined by
\begin{equation}
w = \prod_{i=x,y}
{\rm exp}(\frac{-(E_i^{{\rm miss,calc}} - E_i^{{\rm miss,obs}})^2}{2
  (\sigma_{E_{\rm T}^{\rm miss}})^2}) \\,
\label{eq:nuweight}
\end{equation}
where $E_{x,y}^{\rm{miss,calc}}$ ($E_{x,y}^{\rm{miss,obs}}$) is the
calculated (observed) missing 
transverse momentum component in the $x$ or $y$ direction. 
Solutions that fit better to the expected \ttbar\ event kinematics are
assigned a higher weight. 
The measured resolution of the missing transverse momentum
$\sigma_{E_{\rm T}^{\rm miss}}$ is taken from
Ref.~\cite{METcalibration} as a function of the sum of the transverse
energy in the event. For example, for an event with a total sum of
transverse momentum of $100$~GeV, the resolution is taken to be $6.6$~GeV. 
The weights of all solutions define a weight
distribution for each observable per event. For each event, the
weighted mean value of the respective observable is 
used for the measurement. 

\begin{figure*}[htpb!]
\begin{center}
\subfigure[]{\includegraphics[width=0.49\textwidth]{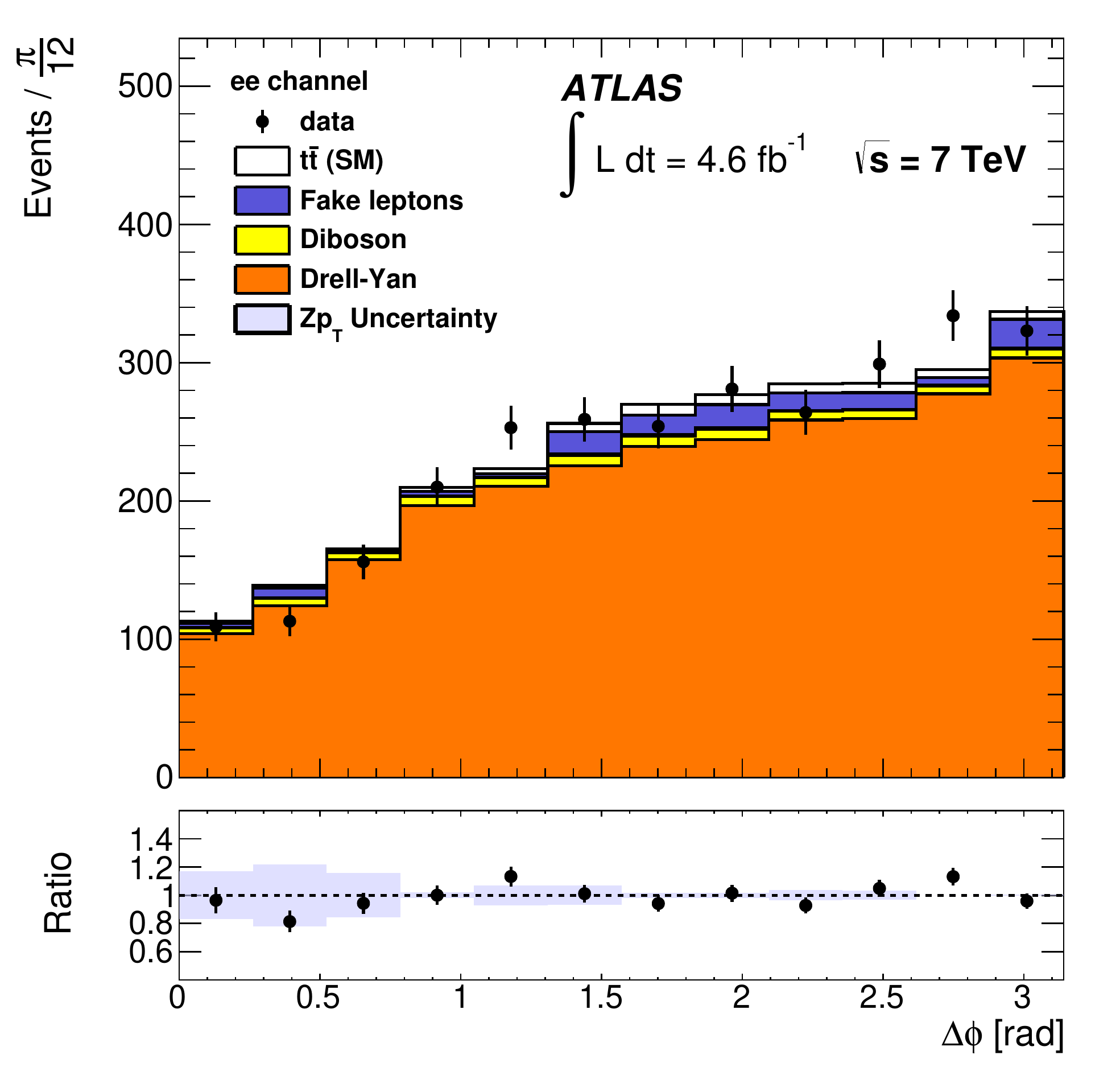}}
\subfigure[]{\includegraphics[width=0.49\textwidth]{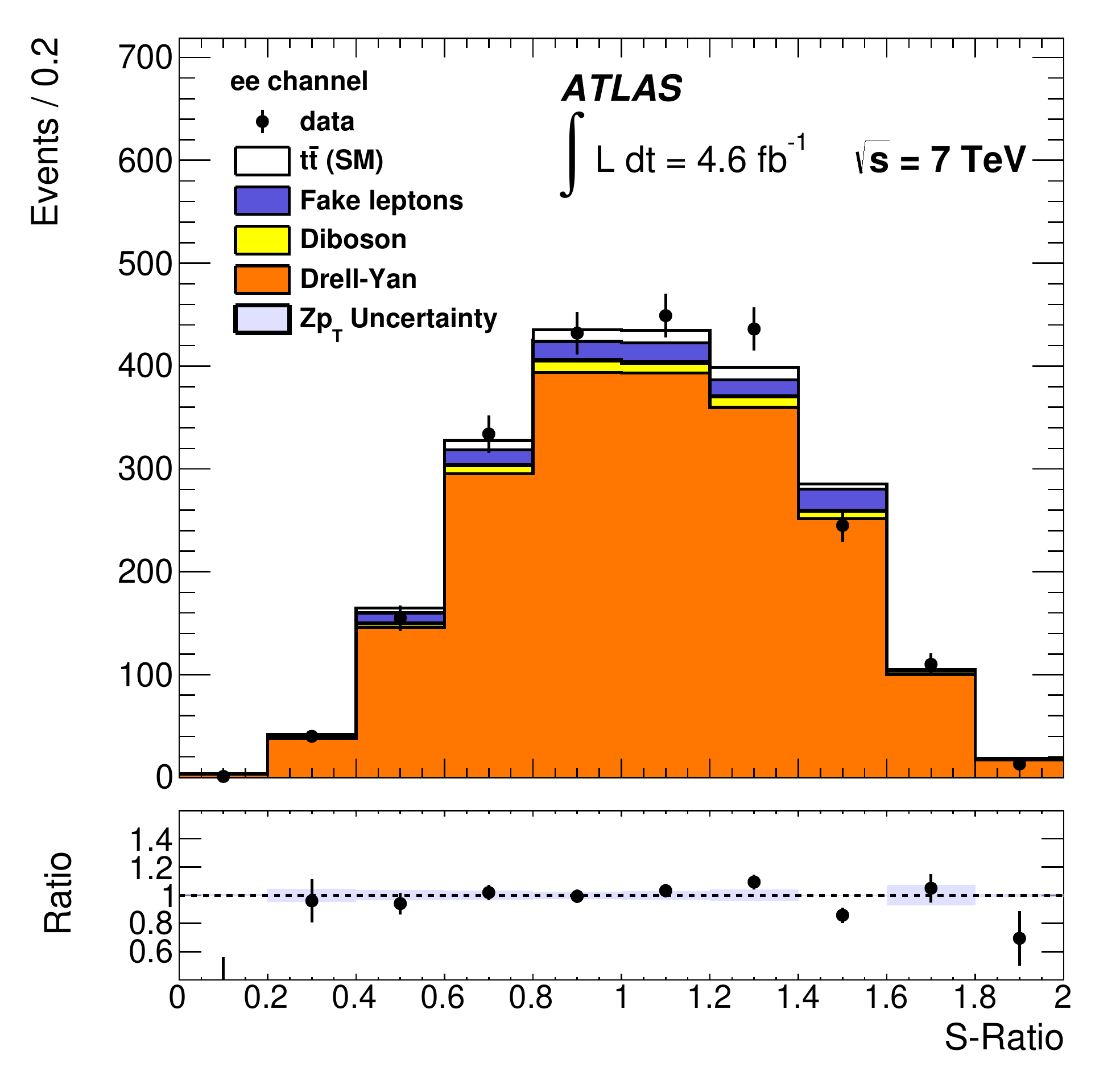}} \\
\subfigure[]{\includegraphics[width=0.49\textwidth]{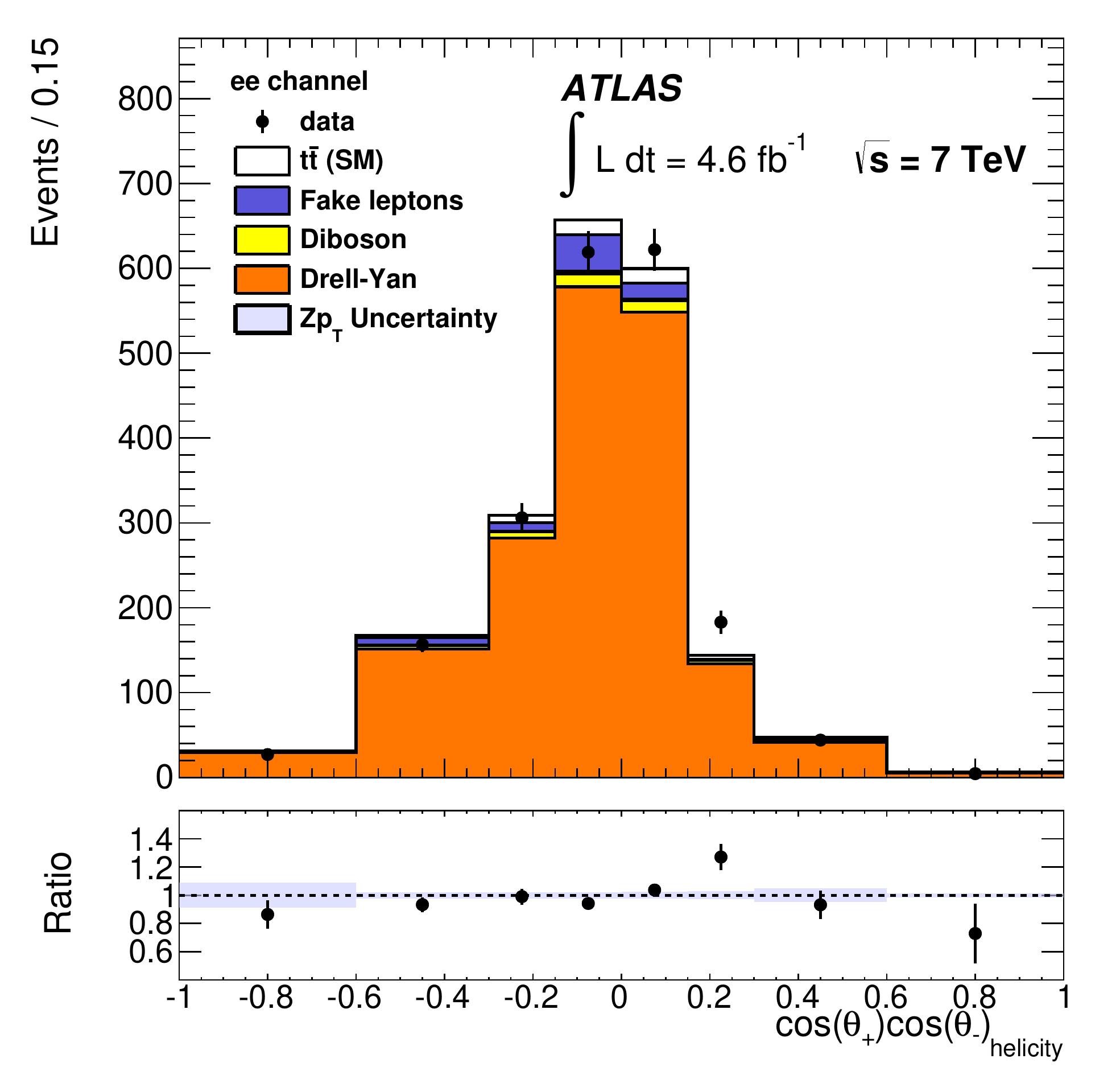}}
\subfigure[]{\includegraphics[width=0.49\textwidth]{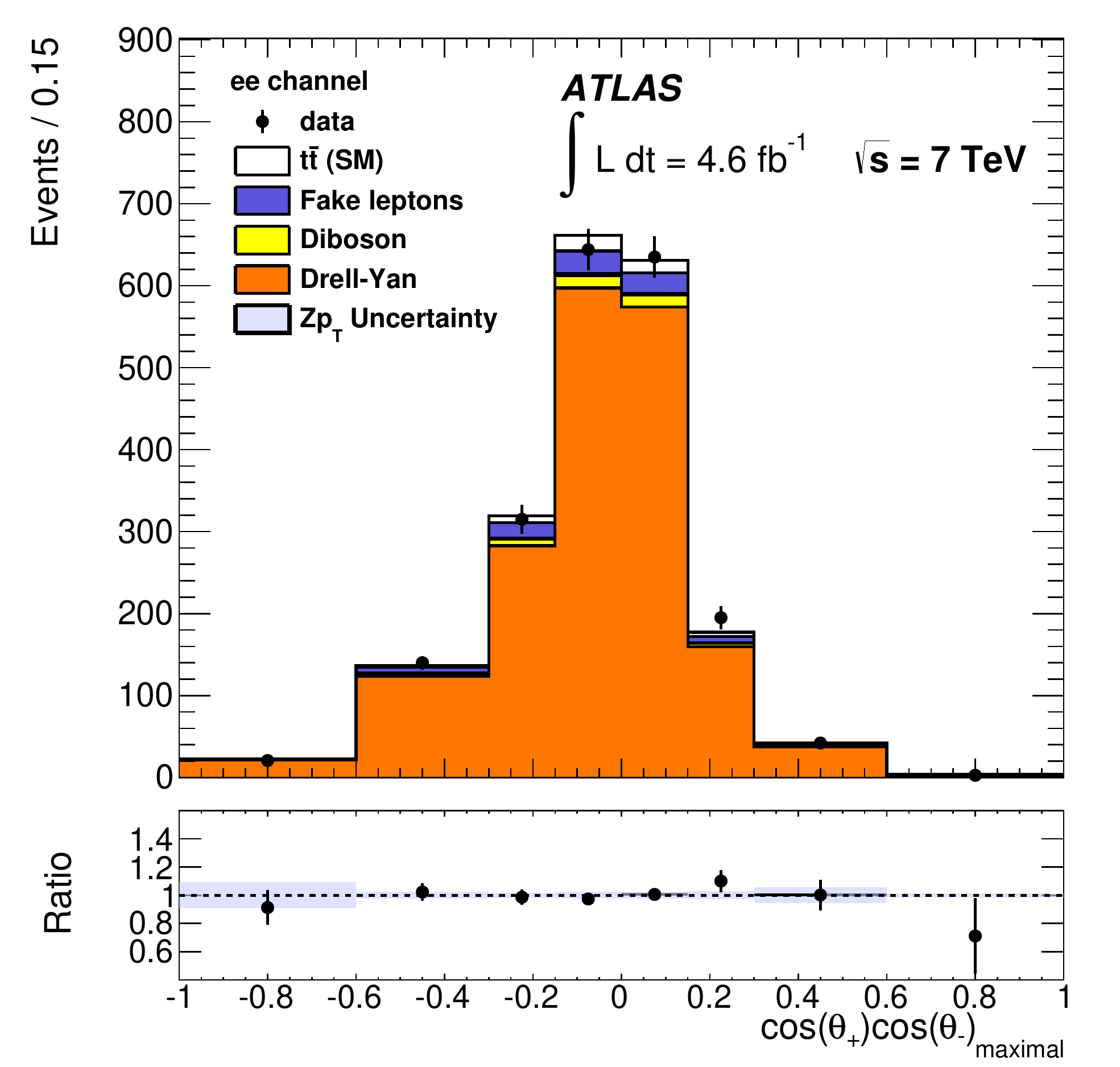}} \\
\caption{\small {Distributions of observables sensitive to \ttbar\ spin
    correlation in the \ee channel in a control region dominated by
    $Z/\gamma^*$+jets background: (a) the azimuthal angle $\Delta \phi(\ell,\ell)$
    between the two charged leptons, (b) the $S$-ratio, as
    defined in Eq.~(\ref{eq:sratio}), (c) $\cos(\theta_+)
    \cos(\theta_-)$, as defined in Eq.~(\ref{eq:coscos}) in the helicity
    basis, and (d) in the maximal basis. The
    $Z/\gamma^*$+jets background is normalized to the data in the
    control region. The contributions from single top and
    $Z\rightarrow\tau^+\tau^-$+jets are not included in the legend as
    their contribution in this region is negligible. The uncertainties
    shown in the ratio are the systematic uncertainty due to the
    modeling of the $Z$ transverse momentum, which is a dominant effect
    in this control region.}}
\label{fig:ee_DY}
\end{center}
\end{figure*}

\begin{figure*}[htpb!]
\begin{center}
\subfigure[]{\includegraphics[width=0.49\textwidth]{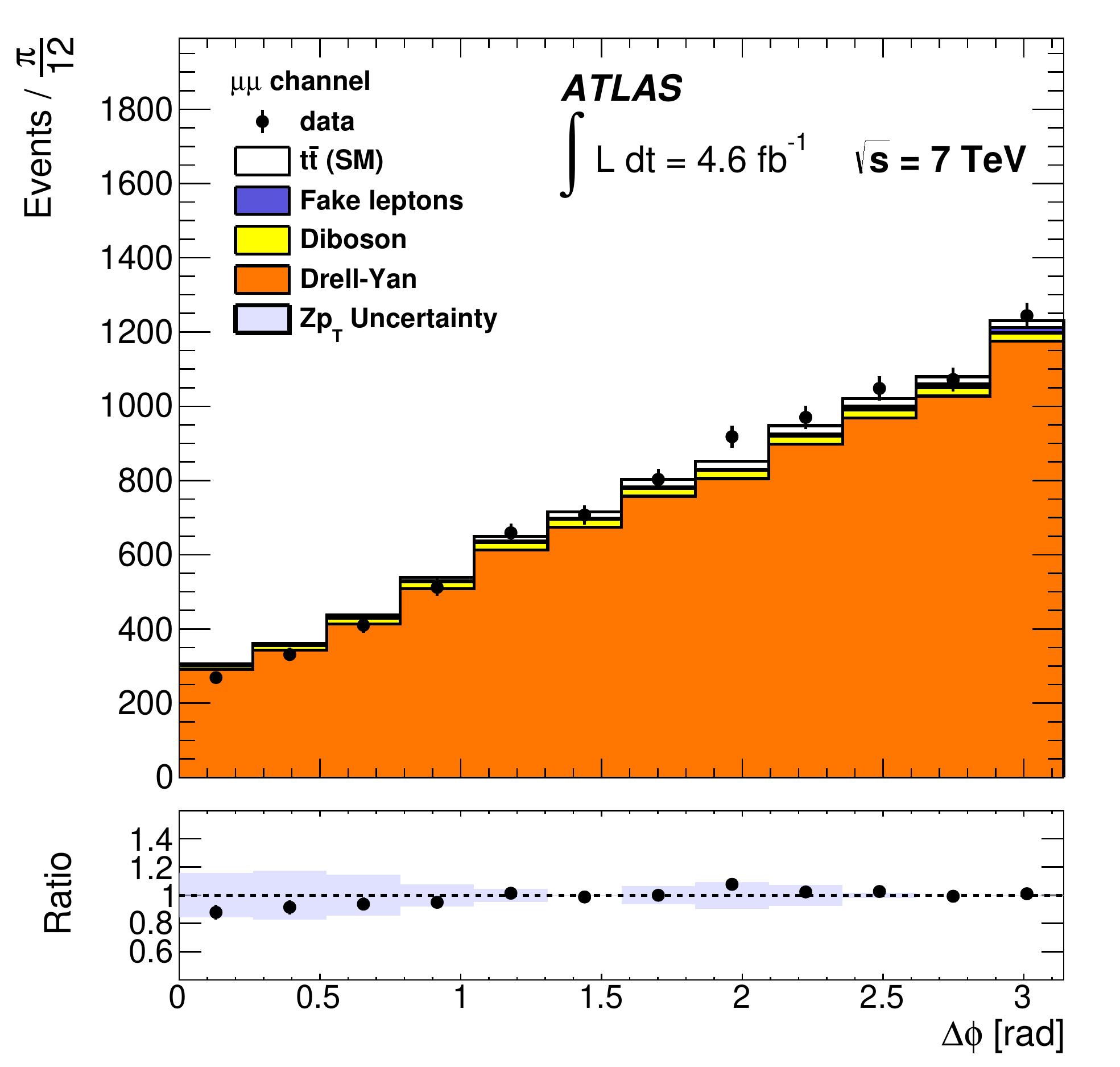}}  
\subfigure[]{\includegraphics[width=0.49\textwidth]{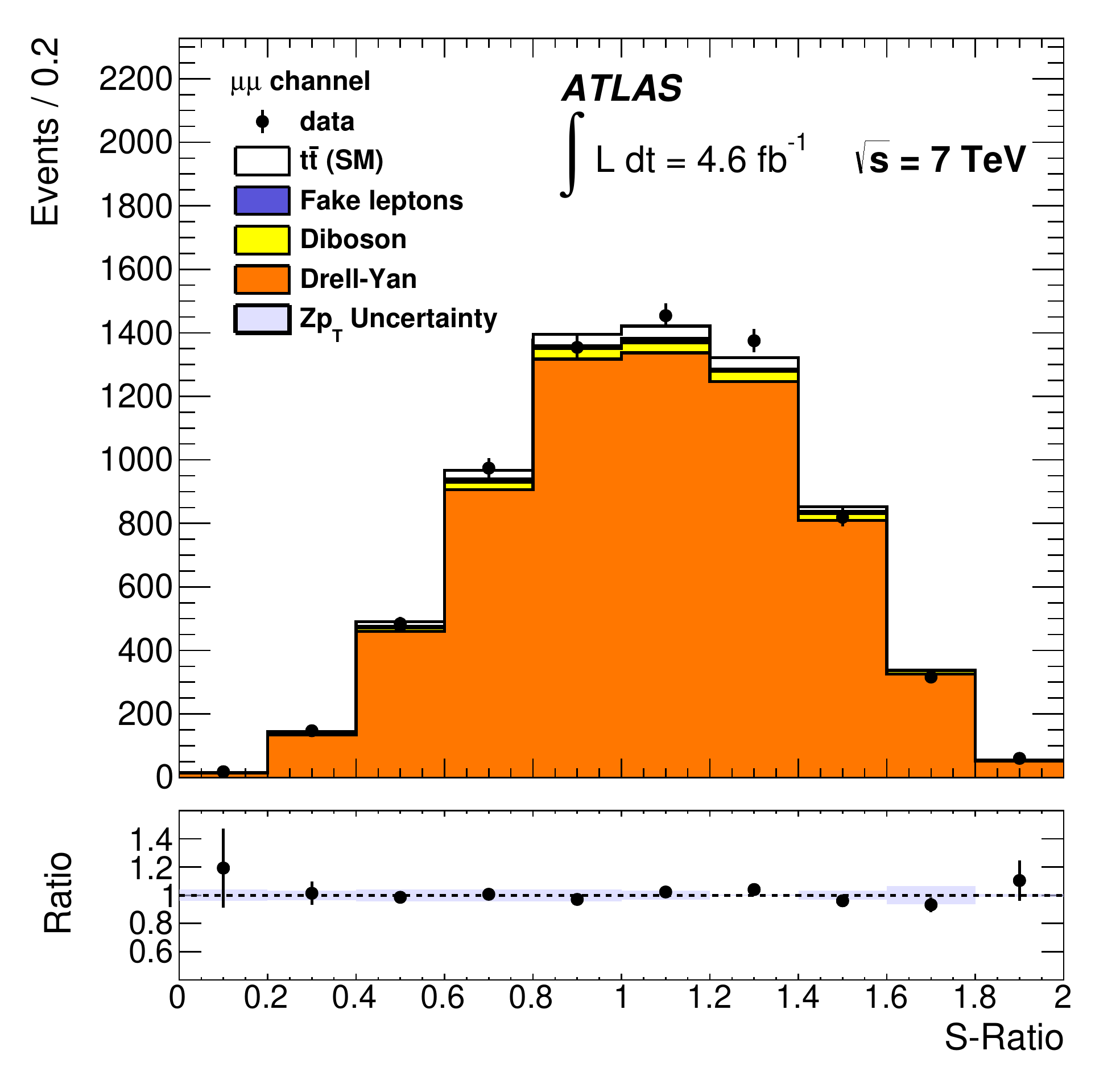}} \\
\subfigure[]{\includegraphics[width=0.49\textwidth]{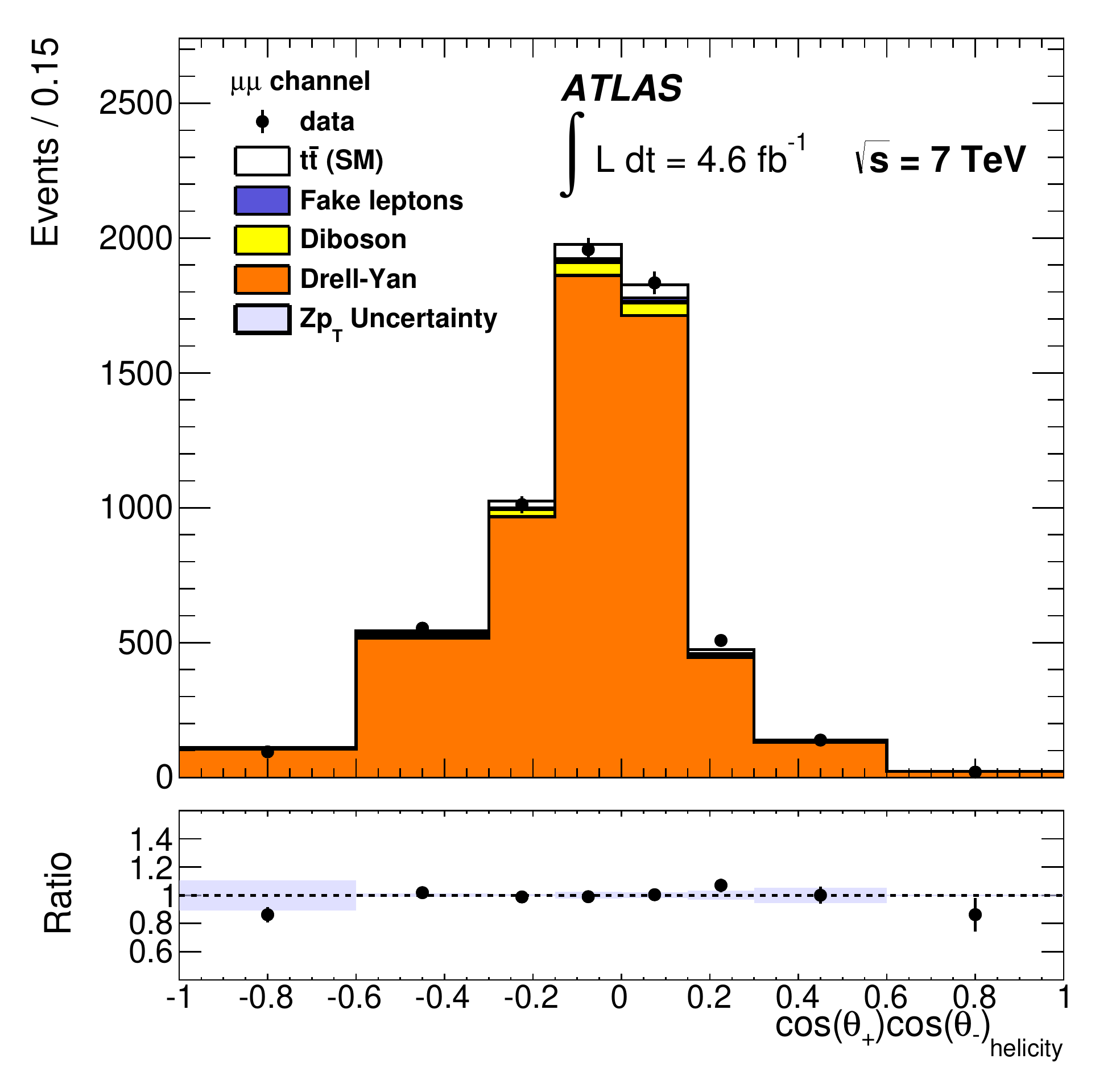}}
\subfigure[]{\includegraphics[width=0.49\textwidth]{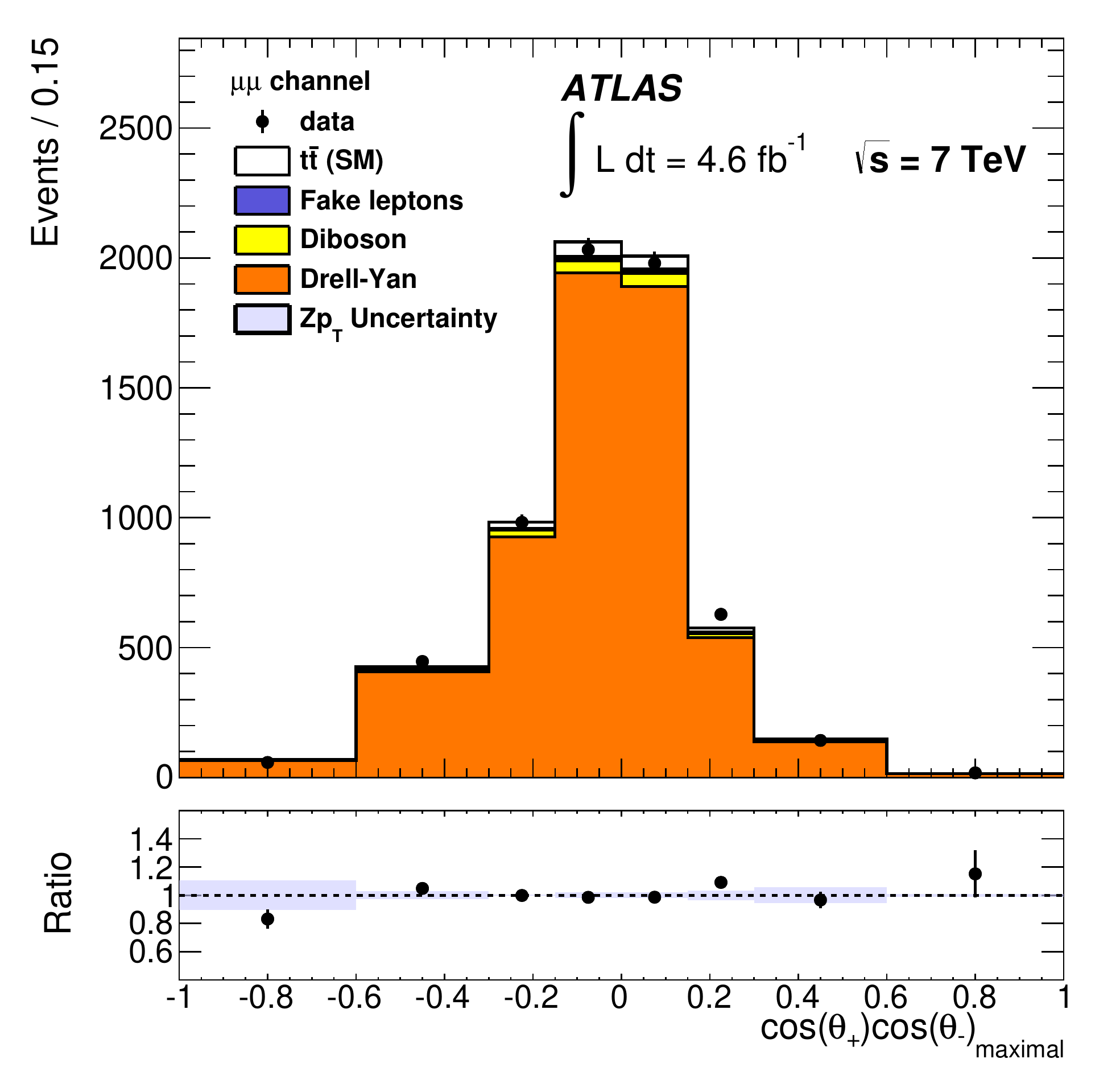}} \\
\caption{\small {Distributions of observables sensitive to \ttbar\ spin
    correlation in the \mumu channel in a $Z/\gamma^*$+jets background
    dominated control region: (a) the azimuthal angle $\Delta \phi(\ell,\ell)$
    between the two charged leptons, (b) the $S$-ratio, as
    defined in Eq.~(\ref{eq:sratio}), (c) $\cos(\theta_+)
    \cos(\theta_-)$, as defined in Eq.~(\ref{eq:coscos}) in the helicity
    basis, and (d) in the maximal basis. The
    $Z/\gamma^*$+jets background is normalized to the data in the
    control region. The contributions from single top and
    $Z\rightarrow\tau^+\tau^-$+jets are not included in the legend as
    their contribution in this region is negligible. The uncertainties
    shown in the ratio are the systematic uncertainty due to the
    modeling of the $Z$ transverse momentum, which is a dominant effect
    in this control region.}}
\label{fig:mumu_DY}
\end{center}
\end{figure*}

Figures~\ref{fig:ee_DY}(b)--\ref{fig:ee_DY}(d)
and~\ref{fig:mumu_DY}(b)--\ref{fig:mumu_DY}(d) show distributions 
of spin correlation observables that use the \ttbar\ event reconstruction with the
neutrino weighting method. For the \ee\ and \mumu\ channels, in a
control region dominated by $Z/\gamma^*$+jets production, the
$S$-ratio and $\cos (\theta_{+}) \cos (\theta_{-})$ in two different spin
quantization bases are presented. 
Good agreement between data and the prediction is observed
confirming a reliable 
description of observables sensitive to \ttbar\ spin correlations with
and without \ttbar\ event reconstruction in the
Drell--Yan background.     

\subsection{Kinematic reconstruction of the \ttbar\ system in the single-lepton
channel} \label{sec:ljetskin}
In the single-lepton events, there is one missing neutrino from the $W \to \ell\nu$ 
decay. Therefore, the $W$-boson mass and the top quark mass can be used 
as constraints to solve the kinematics and to assign the reconstructed
objects (jets, leptons and \met) to the corresponding partons (quarks,
leptons and the neutrino). The main
challenge for the event reconstruction in this final state is
the presence of at least four jets, providing a large number of
possible permutations when assigning objects to partons.

To perform the kinematic reconstruction, the Kinematic Likelihood
Fitter (KLFitter) algorithm~\cite{klfitter} is applied. The likelihood
function is defined as a product of individual likelihood terms 
describing the kinematics of the \ttbar\ signature including constraints
from the masses of the two $W$ bosons and the two top quarks.
Detector resolutions for energy 
measurements are described in terms of transfer functions that map initial
parton kinematics to those of reconstructed jets and leptons.  
The transfer functions are derived for electrons, muons, 
light-quark ($u, d, s, c$) jets and $b$-quark jets,
using a simulated \ttbar\ sample generated with {\sc MC@NLO},
and are parametrized in $\pt$ (for muons) or energy in several $\eta$-regions 
of the detector. The angular variables of
each reconstructed object are measured with a negligible
uncertainty. 

\begin{figure}[htpb!]
\begin{center}
{\includegraphics[width=0.49\textwidth]{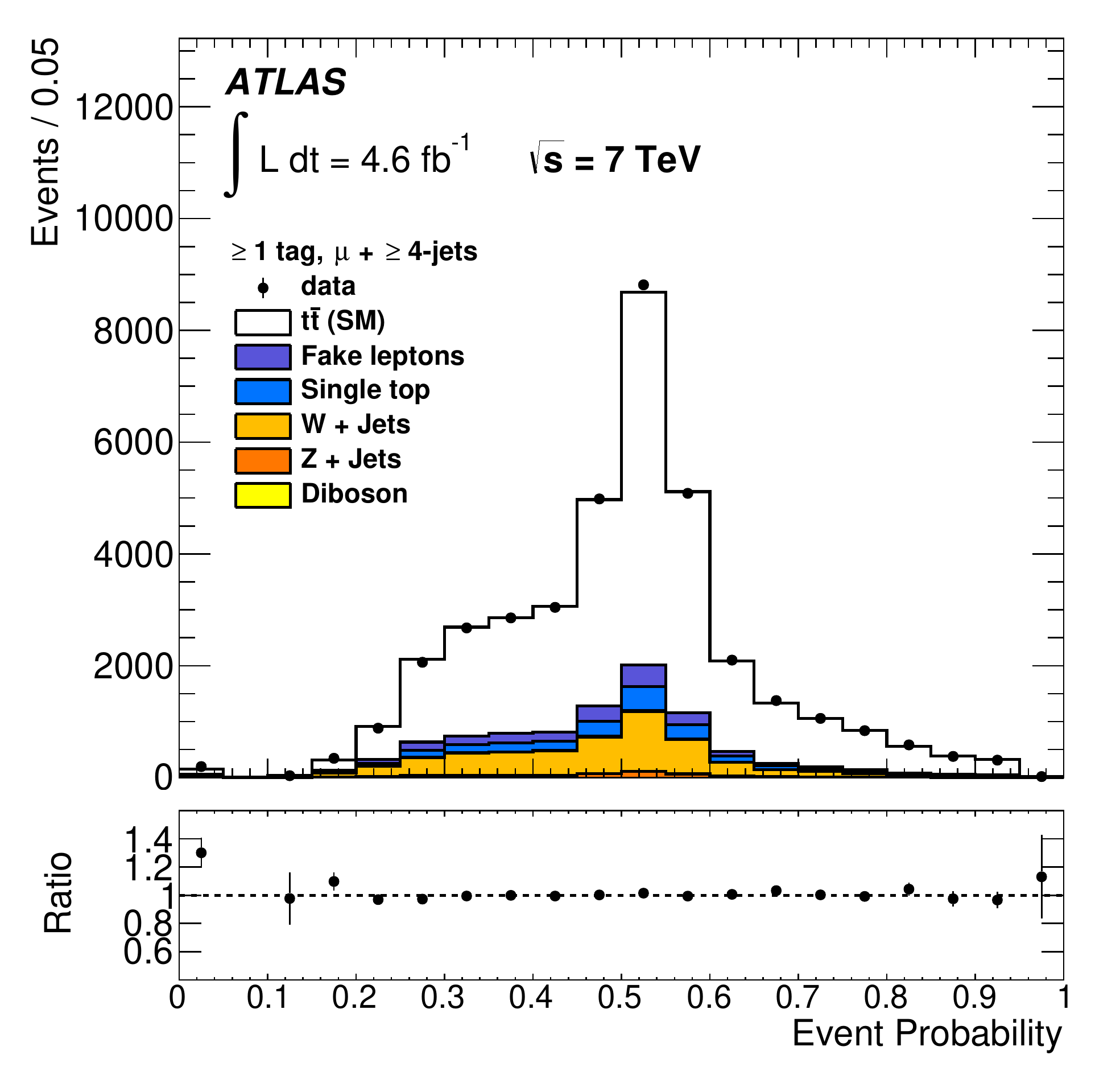}}  
\caption{\small {Event probability distribution in the
    \mujets\ channel for the most likely permutation.}} 
\label{fig:EvProb}
\end{center}
\end{figure}

The likelihood is maximized taking
into account all possible permutations of the objects. 
The maximized likelihood of each permutation is extended to a normalized event probability 
by adding information from $b$-jet identification. This enhances
the probability to choose the correct assignment of the reconstructed objects. 
The likelihood itself is invariant under the 
exchange of jets from down-type and up-type quarks from the
$W$-boson decay. 
To enhance the probability to correctly assign the jets to  down-type
and up-type quarks from the $W$-boson decay, two additional quantities
are incorporated into the likelihood. The first quantity is the
weight assigned to the jet by the $b$-jet tagging algorithm.  This
takes advantage of the fact that 50\% of the $W$-boson decays contain
charm quarks, which have higher $b$-tag weights than other light
quarks. The second quantity is the reconstructed jet \pt. Because of the
$V-A$ structure of the $W$-boson decay, down-type jets have on average
a lower \pt\ than up-type jets. A two-dimensional 
probability distribution of the reconstructed jet \pt\ and the weight
assigned to a jet by the $b$-jet tagging algorithm are used in the event probability.
Figure \ref{fig:EvProb} shows the event probability distribution for
the permutation with the highest value in the \mujets\ channel.

If the \pt\ and $b$-tagging weights of the two light jets are similar,
no additional separation power is obtained and both permutations have
an equal event probability of not larger than 0.5. In case the event
probability reaches values above 0.5, one permutation matches the
model better than all others, implying additional separation power
between the two light jets. 
For the construction of the $\Delta \phi (\ell,d) $ and $\Delta \phi(\ell,b)$ 
observables, the permutation with the highest event probability is chosen. 

\begin{figure*}[htpb!]
\begin{center}
\subfigure[]{\includegraphics[width=0.49\textwidth]{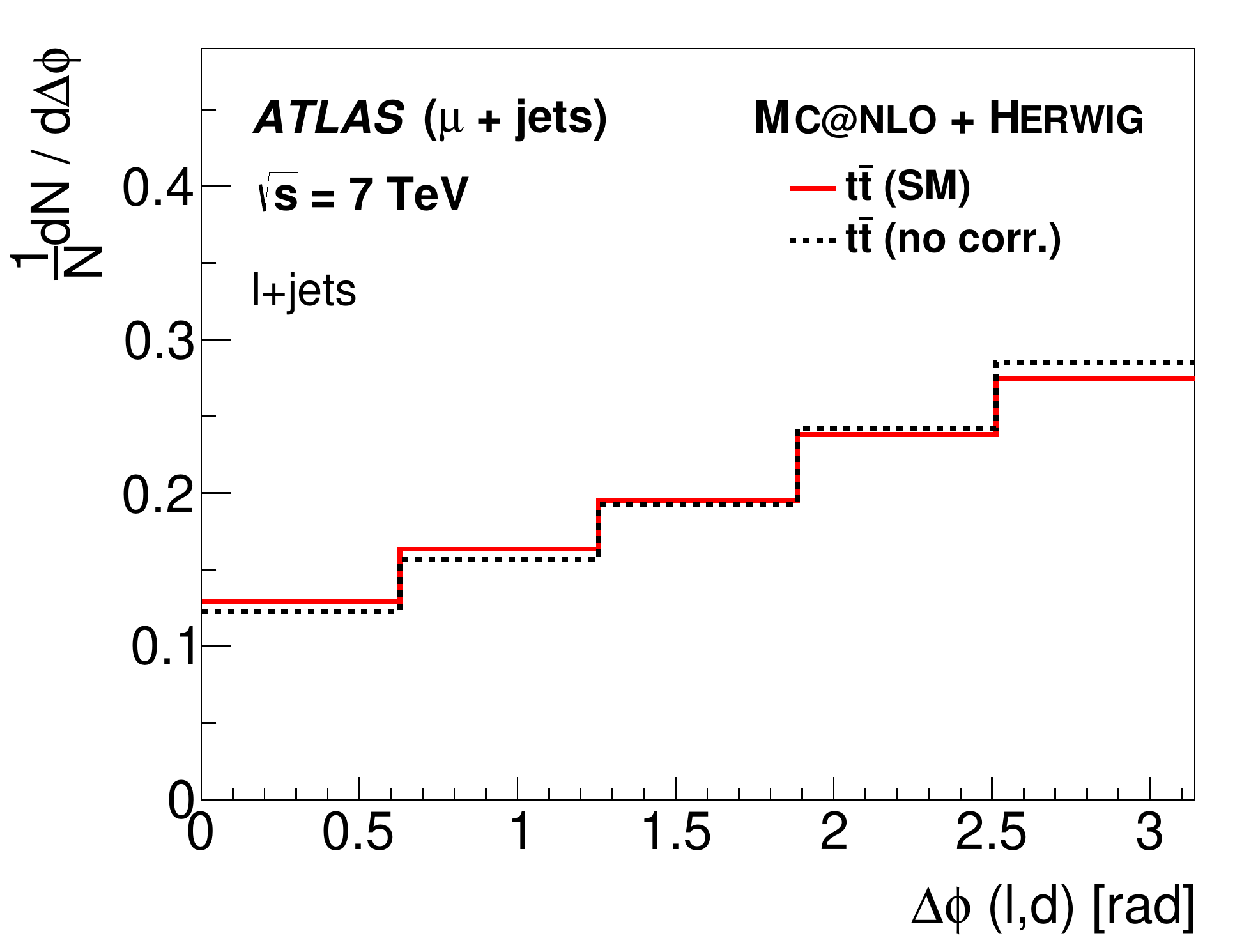}}  
\subfigure[]{\includegraphics[width=0.49\textwidth]{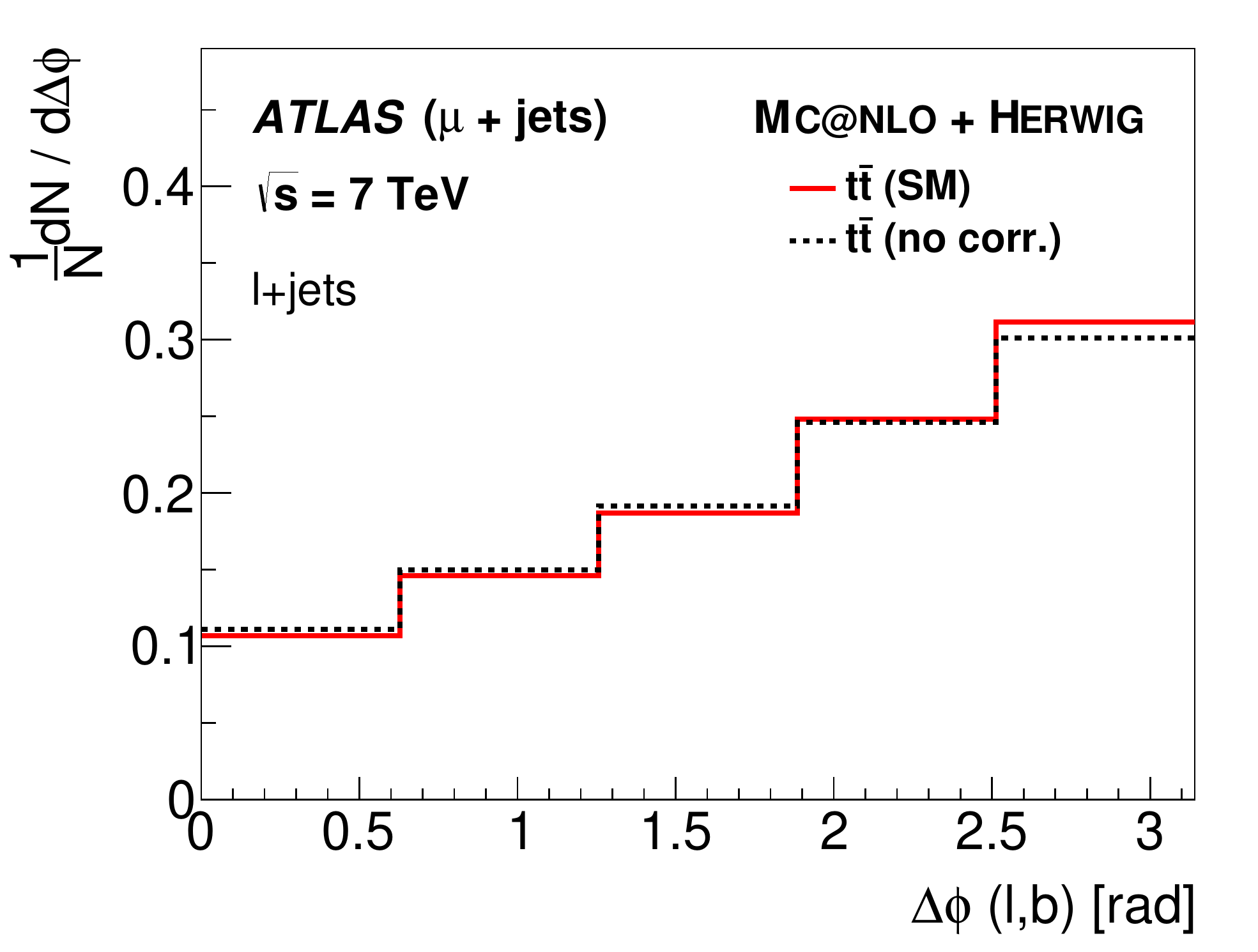}} 
\caption{\small {Distributions of $\Delta \phi (\ell,d)$ between (a) the
    lepton and the jet from the down-type quark
and (b) $\Delta \phi (\ell,b)$ between the lepton and the jet from the
$b$-quark after event selection and reconstruction 
for \mcatnlo\ samples with SM spin correlation and no spin correlation.}}
\label{fig:dphi_reco_wpythia}
\end{center}
\end{figure*}

Figure~\ref{fig:dphi_reco_wpythia} shows distributions 
of $\Delta \phi (\ell,d)$ and $\Delta \phi (\ell,b)$ 
after selection and \ttbar\ kinematic reconstruction for the SM spin correlation 
and no spin correlation scenarios in a sub-channel of single-lepton
events containing one muon and five jets, two of which are $b$-tagged. One can see a 
significant deterioration of the separation between the two
distributions compared
to the parton-level results in Fig.~\ref{fig:obs_parton_lj}. 
This is mainly due to misreconstruction of the top quarks which leads
to a loss of the spin information. 
Because of a more reliable identification of $b$-quark jets compared to 
$d$-quark jets, the separation becomes comparable between the \ddq\ and \dbq\
observables in the single-lepton final state, motivating the use of
both observables for the measurement.    
 
\subsection{Extraction of spin correlation}
\label{sec:extractspin}
To extract the spin correlation strength from the
distributions of the respective observables in data, templates are constructed 
and a binned maximum likelihood fit is performed. 
For each background contribution, one template for every
observable is constructed. For the $t\bar{t}$ signal, one template is constructed 
from a \mcatnlo\ sample with SM spin correlation and another  
using \mcatnlo\ without spin correlation. The templates are fitted to
the data. The  predicted number of events per template bin $i$ is
written as a function 
of the coefficient $f_{\rm SM}$ as 
\begin{equation}
m^{i} = f_{\rm SM} \times m^{i}_{A={\rm SM}}(\sigma_{t\bar{t}}) + (1-f_{\rm SM}) \times m^{i}_{A=0} (\sigma_{t\bar{t}}) + \sum_{j=1}^{N_{\rm bkg}} m^{i}_{j}
\label{eq:nevents}
\end{equation}
where $m^{i}_{A={\rm SM}}(\sigma_{t\bar{t}})$ and $m^{i}_{ A=0}(\sigma_{t\bar{t}})$ 
is the predicted number of signal events in bin $i$ 
for the signal template obtained with the SM \mcatnlo\ sample and 
with the \mcatnlo\ sample with spin correlation turned off, respectively, 
and $\sum_{j=1}^{N_{\rm bkg}} m^{i}_{j}$ is the sum over all
background contributions $N_{\rm bkg}$. 
To reduce the influence of systematic uncertainties sensitive to the
normalization of the signal, the $t\bar{t}$ cross section
$\sigma_{t\bar{t}}$ is included as a free parameter in the fit.

The negative logarithm of the likelihood function $L$
\begin{equation}
L = \prod_{i=1}^{N} {\cal P}(n^{i}, m^{i}) \,
\label{eq:mlikeli}
\end{equation}
is minimized with ${\cal P}(n^{i},m^{i})$ representing the Poisson
probability to observe $n^{i}$ events in bin $i$ with $m^{i}$ events expected.
The number of bins $N$ used for the fit depends on the variable 
and the channel. 

To maximize sensitivity in the single-lepton channel by taking advantage 
of different \ttbar\ signal purities,  the preselected 
sample is split into subsamples of different lepton flavors with exactly 
one and more than one $b$-tagged 
jet and exactly four and at least five jets, thus giving eight subchannels 
in the likelihood fit. Moreover, since the power of the two 
variables \dbq\ and \ddq\ to discriminate between the SM spin
correlation and no spin
correlation scenarios is comparable, and the correlation between them is 
at most 10\%, both are included in the fit as independent subchannels.  
This approach not only allows an effective doubling of the information 
used in the fit but also takes advantage of the opposite behavior of the ratios between 
the spin correlation and no spin correlation scenarios in the two
observables. This in turn leads to   
opposite trends with respect to the signal-modeling systematic
uncertainties resulting in  
significant cancellation effects.  
  
\begin{figure*}[htpb!]
\begin{center}
\subfigure[]{\includegraphics[width=0.49\textwidth]{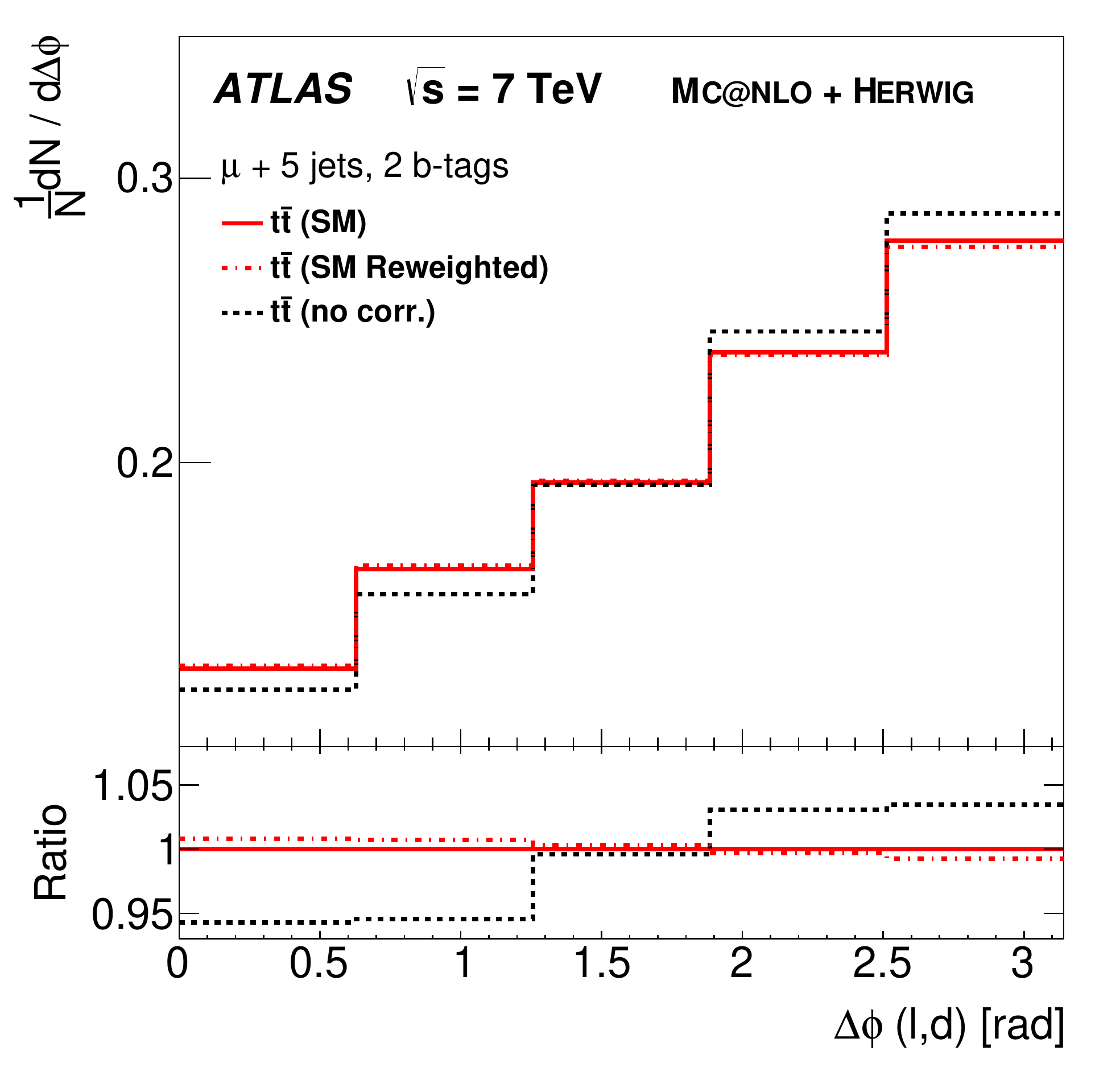}}  
\subfigure[]{\includegraphics[width=0.49\textwidth]{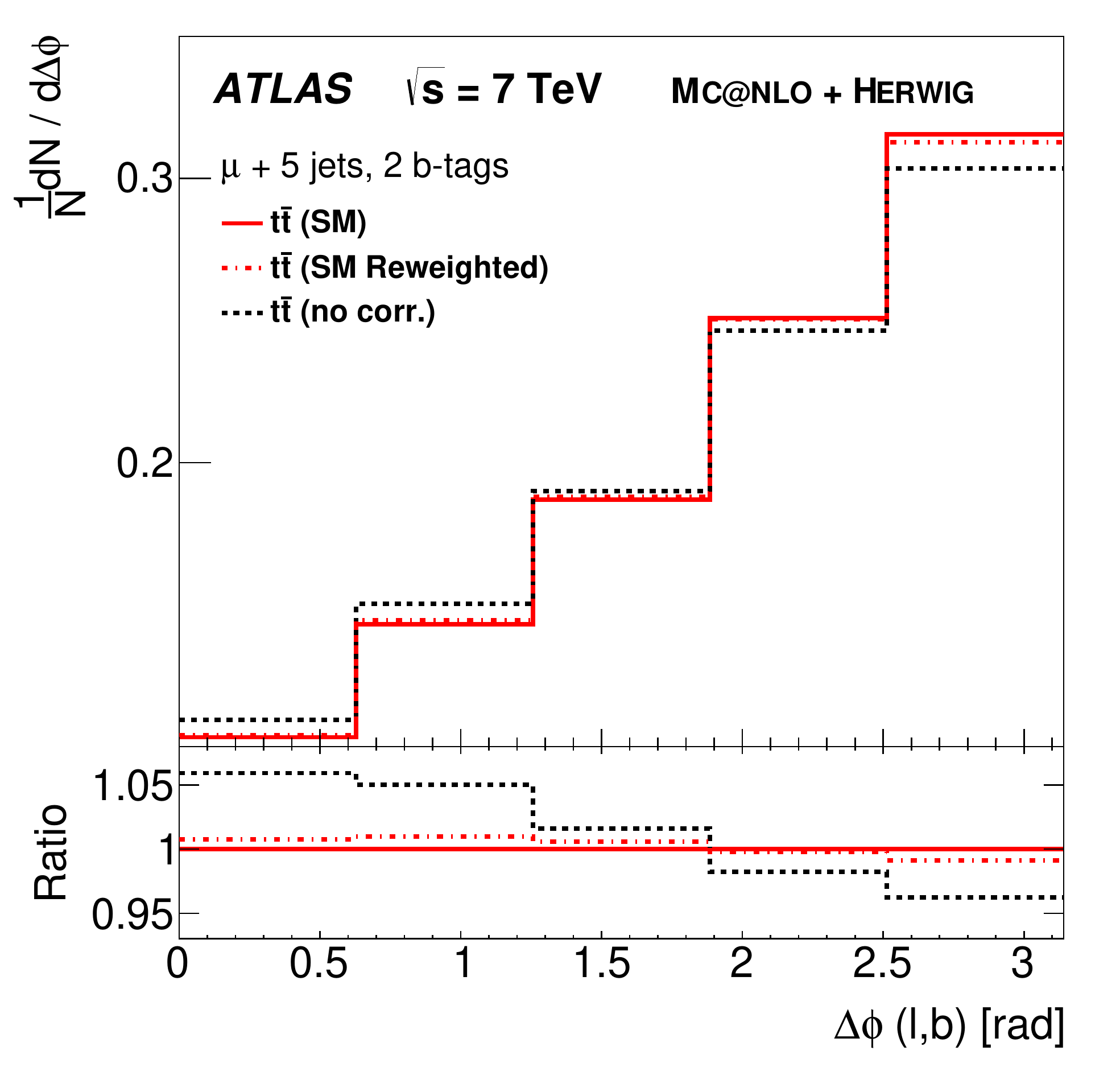}} \\  
\caption{\small {Comparison of the difference of SM spin correlation
    and no spin correlation for (a)
\dbq\ and (b) \ddq\ distributions for the nominal and reweighted-to-{\sc Powheg} 
 top quark \pt\ distributions in the \mcatnlo\ SM spin correlation
 sample. The ``Ratio'' shows the ratio of each distribution to that of
 the SM spin sample.}}
\label{fig:toppt}
\end{center}
\end{figure*}

To demonstrate a reduced sensitivity of the simultaneous fit using  \dbq\ 
and \ddq\ to the choice of the signal model, 
pseudo-data $t\bar{t}$ events simulated
with {\sc Powheg} interfaced to {\sc Herwig} with spin correlation
included ($f_{\rm SM}=1$) were generated and the fit was performed 
using the default templates, simulated with \mcatnlo\ interfaced 
to {\sc Herwig}. The measured 
$f_{\rm SM}$ is $f_{\rm SM} = 1.26 \pm 0.14 {\rm (stat)}$
when using the \ddq\ observable, and $f_{\rm SM} = 0.64
\pm 0.18 {\rm (stat)}$ for \dbq. Fitting both distributions simultaneously resulted
in a value of $f_{\rm SM}$ compatible with the true value, namely
$f_{\rm SM} = 1.02\pm 0.11 {\rm (stat)} $. The difference is explained 
to a large extent by the difference of the top quark \pt\ distributions in 
{\sc Powheg} and \mcatnlo. The recent measurements by the ATLAS~\cite{differ_atlas} 
and CMS~\cite{differ_cms} collaborations indicate that the top quark \pt\ distributions 
vary between the generators and that the top quark \pt\ distribution in data 
is better 
described by {\sc Powheg} interfaced with {\sc Herwig}~\cite{differ_atlas}. 
Ensemble tests performed using templates produced after reweighting the   
top quark \pt\ in the \mcatnlo\ sample to the distribution in {\sc Powheg}  
show a reduced 
difference between the results obtained using different analyzers:  
$f_{\rm SM} = 1.13 \pm 0.14 {\rm (stat)}$ when using 
\ddq, $f_{\rm SM} = 0.77 \pm 0.18 {\rm (stat)}$ for \dbq, and $f_{\rm SM} = 0.99
\pm 0.11 {\rm (stat)}$ if the simultaneous fit to both observables is performed.  
Figures~\ref{fig:toppt}(a) and~\ref{fig:toppt}(b) demonstrate the effect of top quark \pt\ reweighting
on the \ddq\ and \dbq\ distributions, respectively, for the SM spin correlation
sample. One can see that top quark \pt\ reweighting causes the same trend,
but it has the opposite direction with respect to the no spin
correlation and SM spin
correlation hypotheses for the \ddq\ and \dbq\ distributions: for
\ddq\, the reweighting leads to a shape corresponding to larger spin
correlation strength than in the SM, while for 
\dbq\ the shape corresponds to a smaller spin correlation strength.

\section{Systematic Uncertainties}
\label{sec:systematics}
Several classes of systematic uncertainties were considered:
uncertainties related to the detector model and to \ttbar\ signal and background
models. Each source  
can affect the normalization of the signal and the background and/or the shape
of the distributions used to measure the spin correlation
strength. Normalization uncertainties typically have a small effect on
the extracted spin correlation strength since the \ttbar\ cross
section is included as a free parameter in the fit and the contribution of
backgrounds is small.  

Systematic uncertainties are evaluated either by performing pseudo-experiments
or by including them in the fit via nuisance parameters 
represented by Gaussian distributions~\cite{nuisance}. The former is used when no continuous  
behavior of an uncertainty is expected. 
The majority of the uncertainties associated with the modeling 
of signal and background are of non-continuous nature and fall into this category. 
Uncertainties associated with the modeling of reconstruction, 
identification, and calibration of all physics objects used in the analysis  
are included in the fit in the single-lepton 
channel, allowing data to constrain some important uncertainties and thus 
improve sensitivity. In the dilepton channel the effect of the detector 
modeling uncertainties was found to be small and was evaluated by  
performing pseudo-experiments.  

\begin{table*}[htbp]
  \caption{Systematic uncertainties on $f_{\rm SM}$ for the various
    observables in the dilepton final state.
  \label{tab:systematics}}
  \begin{ruledtabular}
  \begin{tabular}{ c c c c c }
  Source of uncertainty                 & $\Delta\phi (\ell,\ell)$  &    $S$-ratio    &    $\cos(\theta_{+}) \cos(\theta_{-})_{\rm helicity}$     &  $\cos(\theta_{+}) \cos(\theta_{-})_{\rm maximal}$ \\
  \hline
  \multicolumn{5}{l}{Detector modeling} \\
  \hline
  Lepton reconstruction                   & \ppm0.01 & \ppm0.02 & \ppm0.05 & \ppm0.03  \\
  Jet energy scale                        & \ppm0.02 & \ppm0.04 & \ppm0.12 & \ppm0.08  \\
  Jet reconstruction                      & $<$0.01  & \ppm0.03 & \ppm0.08 & \ppm0.01  \\
  \met\                                   & \ppm0.01 & \ppm0.01 & \ppm0.03 & \ppm0.02  \\
  Fake leptons                            & \ppm0.03 & \ppm0.03 & \ppm0.06 & \ppm0.04  \\
  \hline
  \multicolumn{5}{l}{Signal and background modeling} \\
  \hline
  Renormalization/factorization scale     &   \ppm0.09  &   \ppm0.08  &   \ppm0.08  &  \ppm0.07   \\
  Parton shower and fragmentation         &   \ppm0.02  &   $<$0.01   &   \ppm0.01  &  \ppm0.08   \\
  ISR/FSR                                 &   \ppm0.08  &   \ppm0.05  &   \ppm0.08  &  \ppm0.01 \\
  Underlying event                        &   \ppm0.04  &   \ppm0.06  &   \ppm0.01  &  $<$0.01   \\
  Color reconnection                      &   \ppm0.01  &   \ppm0.02  &   \ppm0.07  &  \ppm0.07   \\
  PDF uncertainty                         &   \ppm0.05  &   \ppm0.03  &   \ppm0.03  &  \ppm0.05   \\
  Background                              &   \ppm0.04  &   \ppm0.01  &   \ppm0.02  &  \ppm0.02   \\
  MC statistics                           &   \ppm0.03  &   \ppm0.03  &   \ppm0.08  &  \ppm0.04   \\
  Top $p_{\rm T}$ reweighting              &   \ppm0.09  &   \ppm0.03  &   \ppm0.03  &  $<$0.01    \\
  \hline                                                                                       
  Total systematic uncertainty            &   \ppm0.18  &   \ppm0.14  &   \ppm0.23 &   \ppm0.18   \\
  Data statistics                         &   \ppm0.09  &   \ppm0.11  &   \ppm0.19  &  \ppm0.14   \\
  \end{tabular}
  \end{ruledtabular}
\end{table*}

Pseudo-experiments are created according to the following procedure. 
For each source of uncertainty templates corresponding to the 
respective up and 
down variation are created for both the SM and the uncorrelated spin templates, 
taking into account the change of the acceptance and shape 
of the observable due to the source under study.
Pseudo-data sets are generated by mixing these templates according to the 
measured $f_{\rm SM}$ and applying Poisson fluctuations to each bin. Then the  
nominal and varied templates are used to perform a fit to the same 
pseudo-data. This procedure is repeated many times
for each source of systematic uncertainty, 
and the means of the differences between the central fit values and
the up and down variations are symmetrized and quoted as the
systematic uncertainty from this source. Systematic uncertainties arising 
from the same source are treated as correlated between different dilepton 
or single-lepton channels.   

Uncertainties in the detector model include uncertainties 
associated with the objects used in the event selection. Lepton uncertainties 
(quoted as ``Lepton reconstruction'' in Table~\ref{tab:systematics}) include 
trigger efficiency and identification uncertainties for electrons and
muons, and  
uncertainties due to electron 
(muon) energy (momentum) calibration and resolution. 
Uncertainty associated with the jet energy calibration is referred to as 
``Jet energy scale'', while jet reconstruction efficiency and resolution uncertainties 
are combined and quoted as ``Jet reconstruction'' in  Table~\ref{tab:systematics}. 
Uncertainties on the \met\ include uncertainties due to 
the pile-up modeling and the modeling of energy deposits not 
associated with the reconstructed objects.   

A number of systematic uncertainties affecting the $t\bar{t}$ modeling
are considered. Systematic uncertainty associated
with the choice of factorization and renormalization scales in 
\mcatnlo\ is evaluated by varying the default scales by a factor of two 
up and down simultaneously. The uncertainty due to the   
choice of parton shower and hadronization model is determined by 
comparing two alternative samples simulated with the {\sc Powheg (hvq v4)}~\cite{powheg} 
generator interfaced 
with {\sc Pythia 6.425}~\cite{pythia} and {\sc Herwig} v6.520. The uncertainty on  
the amount of initial- and final-state radiation (ISR/FSR) in the
simulated \ttbar\ sample is  
assessed by comparing {\sc Alpgen}, showered with {\sc Pythia}, with varied 
amounts of initial- and final-state radiation. The size of the
variation is compatible with the recent measurements of additional jet 
activity in $t\bar{t}$ events~\cite{vetopaper}. The uncertainty due to the     
choice of the underlying event model is estimated 
by comparing a {\sc Powheg}-generated sample showered by {\sc Pythia} 
with the {\sc Perugia} 2011 tune to one with 
the {\sc Perugia} 2011 {\sc mpiHi} tune~\cite{UEtunes}. The latter is a
variation of the {\sc Pythia} 2011 tune with more semi-hard multiple parton
interactions. The impact of the color reconnection model of the partons 
that enter hadronization is assessed by comparing samples  
generated with {\sc Powheg} and showered by {\sc Pythia} with 
the {\sc Perugia} 2011 tune and the {\sc Perugia} 2011 {\sc noCR} 
tune~\cite{UEtunes}. 
To investigate the effect of the choice of PDF used in the analysis,
the uncertainties from the nominal CT10 PDF set and from the
NNPDF2.3~\cite{nnpdf} and MSTW2008~\cite{mstw} NLO PDF sets are considered and 
the envelope of these uncertainties is taken as the
uncertainty estimate.     
The dependence of the measured $f_{\rm SM}$ on the top quark mass is
evaluated by changing the value of 172.5~\GeV\ used in the simulation and performing a linear fit
of the dependency of the considered observable on the top quark mass within
the mass range $172.5 \pm 5$ \GeV.

\begin{table*}[htbp]
 \caption{Summary of $f_{\rm SM}$ measurements in the individual
   dilepton channels and in the
 combined dilepton channel for the four different observables. The
 uncertainties quoted are first statistical and then systematic.
 \label{tab:results}}
\hspace*{-0.8cm}
\begin{ruledtabular}
\begin{tabular}{c|c@{\hskip 0.3in}ccc}
 Channel  & $f_{\rm SM}(\Delta \phi (\ell,\ell))$ & $f_{\rm SM}(S$-ratio)
 & $f_{\rm SM}(\cos(\theta_{+}) \cos(\theta_{-})_{\rm helicity})$ 
 & $f_{\rm SM}(\cos(\theta_{+}) \cos(\theta_{-})_{\rm maximal})$  \\       
\hline
 \ee\     & $0.87\pm0.35\pm0.50$ & $0.81\pm0.35\pm0.40$   & $1.72\pm0.57\pm0.75$ & $0.48\pm0.41\pm0.52$ \\
 \emu\    & $1.24\pm0.11\pm0.13$ & $0.95\pm0.12\pm0.13$   & $0.76\pm0.23\pm0.25$ & $0.86\pm0.16\pm0.20$ \\
 \mumu\   & $1.11\pm0.20\pm0.22$ & $0.53\pm0.26\pm0.39$   & $0.31\pm0.42\pm0.58$ & $0.97\pm0.33\pm0.44$ \\ 
 \hline                                                                                                  
 Dilepton & $1.19\pm0.09\pm0.18$ & $0.87\pm0.11\pm0.14$   & $0.75\pm0.19\pm0.23$ & $0.83\pm0.14\pm0.18$ \\
 \end{tabular}
 \end{ruledtabular}
 \end{table*}

Uncertainties on the backgrounds (quoted as ``Background'' in 
Table~\ref{tab:systematics}), evaluated using simulation, arise
from the limited knowledge of the theoretical cross sections for
single top, diboson and $Z \rightarrow \tau^{+}\tau^{-}$ production, from
the modeling of additional jets in these samples and from the 
integrated luminosity. The uncertainty of the latter amounts to 
$\pm1.8$\%~\cite{lumi}.   
Systematic uncertainties on the $Z \rightarrow \ee$ and $Z \rightarrow \mumu$  
backgrounds result from the uncertainty of their normalization to data in
control regions and modeling of the $Z$-boson transverse momentum. 
It was checked that these uncertainties cover the small
differences between data and prediction seen in
Figs.~\ref{fig:ee_DY} (a) and~\ref{fig:mumu_DY} (a).  
The uncertainty on the $W$+jets background in the single-lepton channel 
arises from the normalization uncertainty and from the uncertainty 
on the flavor composition given by the charge asymmetry method. 
The uncertainty on the fake lepton background (``Fake leptons'' in 
Table~\ref{tab:systematics}) 
arises mainly from uncertainties on the measurement of lepton
misidentification rates in different control samples. 
 
Finally, an uncertainty on the method to extract the spin correlation
strength arises from the limited size of the MC samples used to create the templates.

As discussed in Sec.~\ref{sec:measurement},  top quark
\pt\ modeling has an effect on $f_{\rm SM}$. The effect on $f_{\rm
  SM}$ of reweighting of the top quark
\pt\ to match the distribution in unfolded data is listed
separately in Sec.~\ref{sec:extractspin}. 
To avoid  double counting, the uncertainty due to the choice of parton
shower and hadronization  
model is evaluated after the top quark \pt\ distribution in {\sc Powheg}+{\sc Pythia} is 
corrected to be consistent with {\sc Powheg}+{\sc Herwig}. 

\section{Results}
\label{sec:results}
In the following, the results for the spin correlation measurements in
the dilepton and single-lepton final states are discussed. 
\subsection{Dilepton channel}
For each of the four  observables, the maximum
likelihood fit in each of the three individual channels (\ee, \emu, and
\mumu) and their combination is performed. 
The observable with the largest statistical separation power between the no spin 
correlation and the SM spin correlation hypotheses is $\Delta \phi$.
The measured values of $f_{\rm SM}$ for $\Delta \phi (\ell,\ell)$, 
the $S$-ratio and
$\cos(\theta_{+}) \cos(\theta_{-})$  
in the helicity and maximal bases are summarized in
Table~\ref{tab:results}. 
The systematic uncertainties and their effect on the measurement of
$f_{\rm SM}$ in the dilepton channel are listed in Table~\ref{tab:systematics}. 
Because of the different methods of constructing the four observables,
they have different sensitivities to the various sources of systematic
uncertainty and to the various physics effects. Some of the given
uncertainties are limited by the size 
of the samples used for their extraction. The dependence of $f_{\rm
  SM}$ on the top quark mass $m_t$ is parametrized as 
$\Delta f_{\rm SM} = -1.55 \times 10^{-5} (m_t/\rm GeV  - 172.5)$
for $\Delta \phi (\ell,\ell)$, $\Delta f_{\rm SM} = - 0.010 (m_t/\rm
GeV - 172.5)$ for the $S$-ratio, $\Delta f_{\rm SM} = 0.015 (m_t/\rm
GeV  - 172.5)$ for $\cos(\theta_{+}) \cos(\theta_{-})$ in the helicity
basis, and  $\Delta f_{\rm SM} = 0.016 (m_t/\rm GeV  - 172.5)$ for
$\cos(\theta_{+}) \cos(\theta_{-})$ in the maximal basis. 

\begin{figure*}[htpb!]
\begin{center}
\subfigure[]{\includegraphics[width=0.49\textwidth]{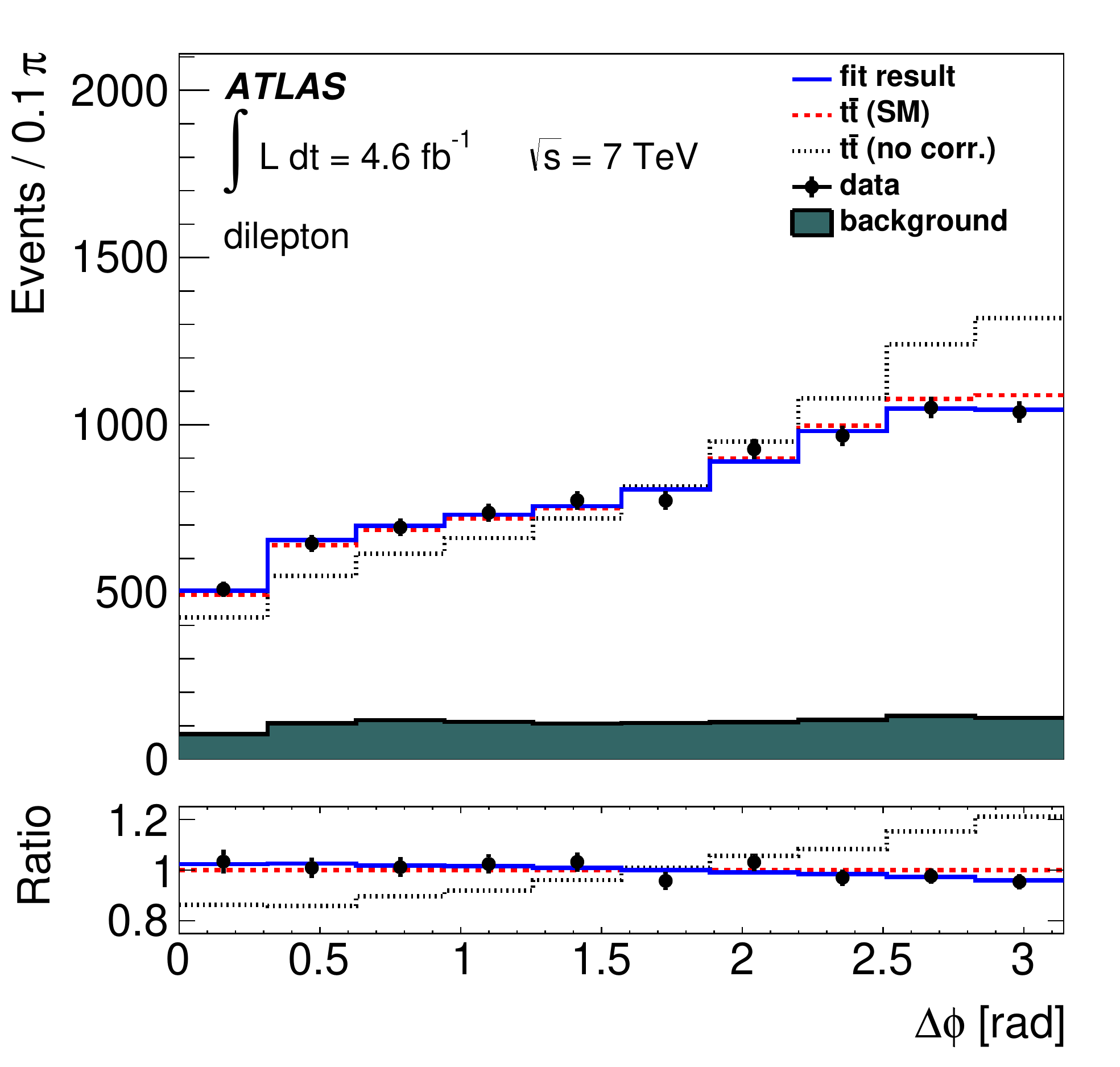}}
\subfigure[]{\includegraphics[width=0.49\textwidth]{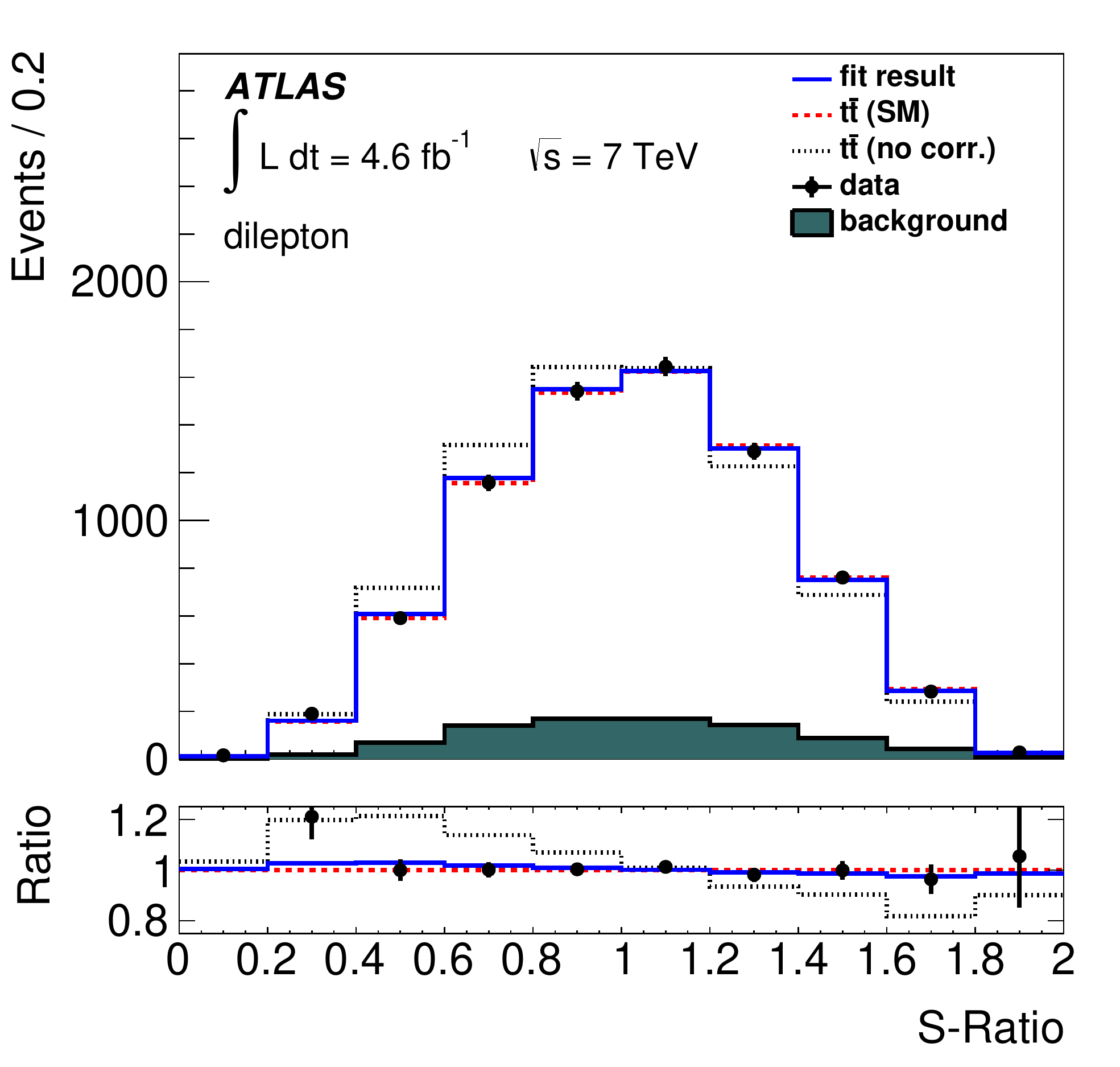}}\\
\subfigure[]{\includegraphics[width=0.49\textwidth]{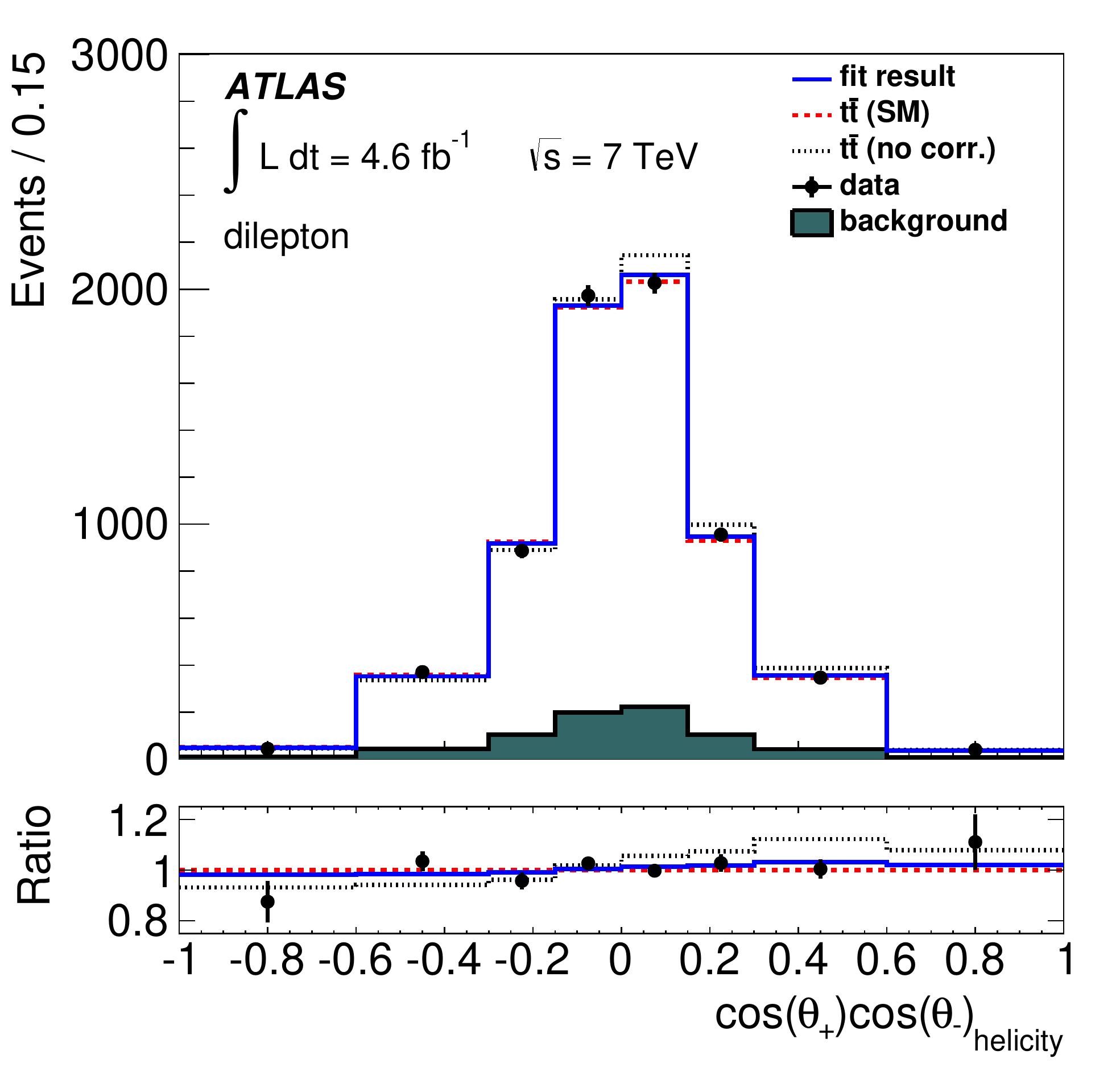}}
\subfigure[]{\includegraphics[width=0.49\textwidth]{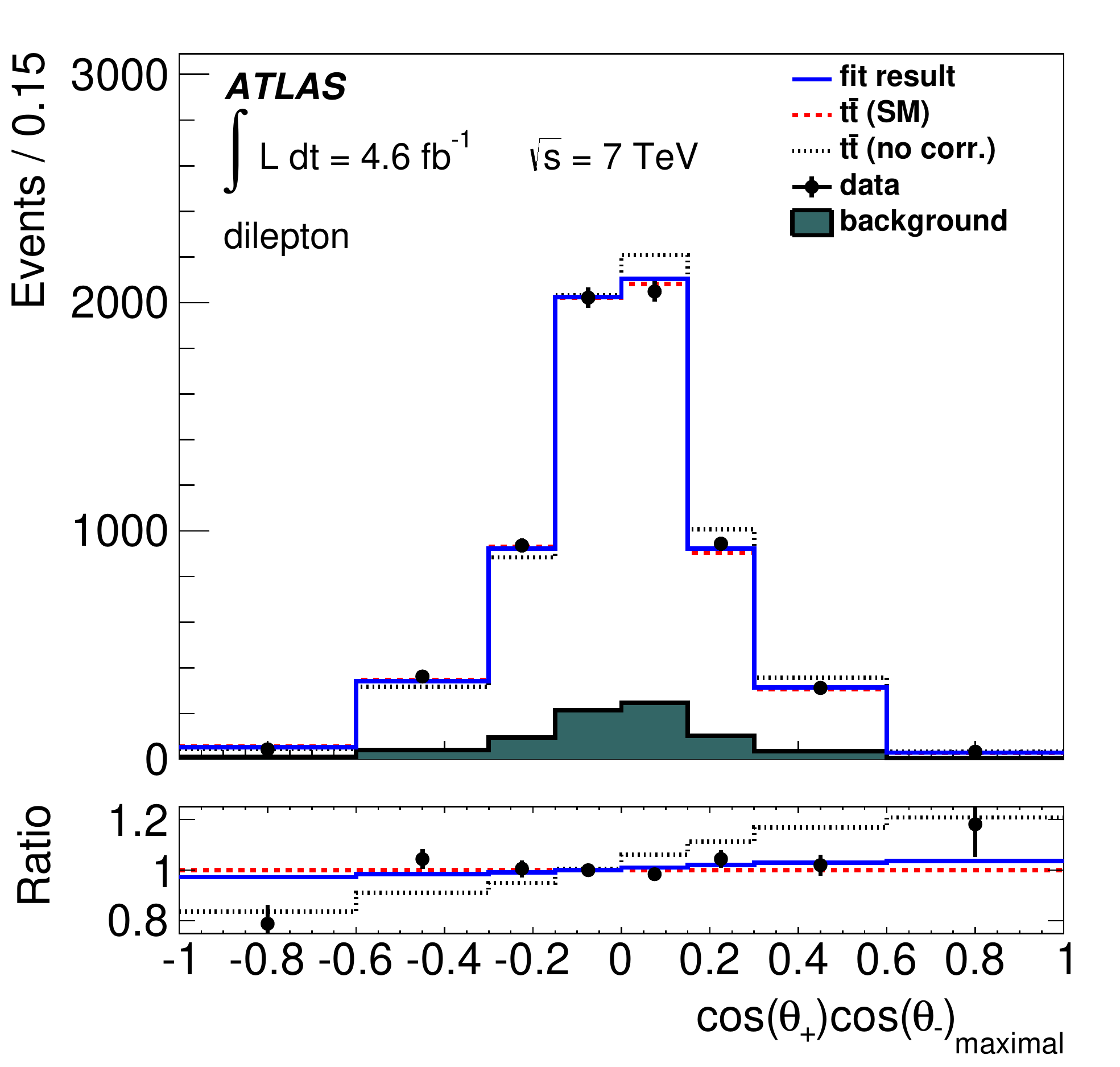}}
\end{center}
\caption{Distributions of (a) $\Delta\phi (\ell,\ell)$, (b) $S$-ratio, (c)
  $\cos(\theta_{+}) \cos(\theta_{-})$ in the helicity basis, and (d)
  $\cos(\theta_{+}) \cos(\theta_{-})$ in the maximal basis in the
  dilepton final state. The result
  of the fit to data (blue lines) is compared 
to the templates for background plus \ttbar\ signal with SM spin
correlation (red dashed lines) and without spin correlation (black
dotted lines). The bottom panel shows the ratio of the data (black
points), the best fit (blue solid lines) and the no spin prediction to
the SM prediction.   
}
\label{fig:finalplots}
\end{figure*}

Figure~\ref{fig:finalplots} shows the distribution of the four
observables in the data, the prediction for SM spin correlation and no
spin correlation, and the result of the fit. 

The analysis of the $\cos(\theta_{+}) \cos(\theta_{-})$ observable allows a direct  
measurement of the spin correlation strength $A$, because $A$ is defined
by the $\cos(\theta_{+}) \cos(\theta_{-})$ distribution according to
Eq.~(\ref{eq:coscos}).  
This becomes obvious in
Eqs.~(\ref{eq:C}) and~(\ref{eq:C_expect}), which show that the expectation value of
$\cos(\theta_{+}) \cos(\theta_{-})$ is equal to $A$ modulo constant factors.
Therefore, the extraction of $f_{\rm SM}$ using the full
distribution in a template method is equivalent to extracting the
spin correlation in the respective spin quantization basis $A_{\rm
  basis}^{\rm measured}$. The relation is given by 
\begin{equation}\label{Eq:fSM_vs_A}
A_{\rm basis}^{\rm measured}= f_{\rm SM} \, A_{\rm basis}^{\rm SM} \, ,
\end{equation}
with the SM predictions being $A_{\rm helicity}^{\rm SM}=0.31$ and $A_{\rm
  maximal}^{\rm SM}=0.44$, respectively, as discussed in
Sec.~\ref{sec:observables}. 

Combining all three final states in the measurement of
$\cos(\theta_{+}) \cos(\theta_{-})$ in the helicity basis, a
direct measurement of 
$A_{\rm helicity}^{\rm measured}= 0.23 \pm 0.06~{\rm (stat)}\pm 0.07~{\rm (syst)}$
is derived, which is
in good agreement with the SM value of $A_{\rm helicity}^{\rm SM}=0.31$.

The combined result using $\cos(\theta_{+}) \cos(\theta_{-})$ in the
maximal basis gives a direct measurement of
$A_{\rm maximal}^{\rm measured}= 0.37 \pm 0.06~{\rm (stat)}\pm0.08~{\rm (syst)}$,
in good agreement with the SM value of $A_{\rm  maximal}^{\rm SM}=0.44$. 

The analysis of $\Delta \phi(\ell,\ell)$ and the $S$-ratio allows an
indirect extraction of $A$ under the assumption that the $t \bar t$ sample
is composed of top quark pairs as predicted by the SM, either with or
without spin correlation, but does not contain contributions beyond
the SM. In that case, a change in the fraction $f_\text{SM}$ will lead 
to a linear change of $A$ according to Eq.~\ref{eq:coscos}. This has
been verified in pseudoexperiments. Under these conditions, the
measured $f_{\rm SM}$ can be 
translated into values of $A_{\rm basis}^{\rm measured}$ via
Eq.~\ref{Eq:fSM_vs_A}, giving $A_{\rm helicity}^{\rm measured} = 0.37 \pm
0.03~{\rm (stat)} \pm 0.06~{\rm ( syst)}$ and
$A_{\rm maximal}^{\rm measured} = 0.52 \pm 0.04~{\rm (stat)} \pm
0.08~{\rm (syst)}$.
These results are limited by
systematic uncertainties, in particular by uncertainties due to signal
modeling. The influence of the dominant systematic
uncertainties in the previous ATLAS measurement performed on a smaller
data set ($2.1$~fb$^{-1}$), giving $A_{\rm 
  helicity}=0.40^{+0.09}_{-0.08} ~({\rm stat} \oplus {\rm syst})$~\cite{atlasspin},
has been reduced due to a better model of the fake lepton background and improved
understanding of the jet energy scale. The two results are in
agreement with each other.

The analysis of the $S$-ratio results in
$A_{\rm helicity}^{\rm measured}  =   0.27 \pm 0.03~{\rm (stat)}\pm0.04~{\rm (syst)}$ and
$A_{\rm maximal}^{\rm measured}  =  0.38 \pm 0.05~{\rm (stat)}\pm 0.06~{\rm(syst)}$. 

\begin{table*}[htbp]
 \caption{Summary of measurements of the spin correlation strength $A$
   in the helicity and maximal bases in the
 combined dilepton channel for the four different observables. For the
 indirect extractions using $\Delta \phi(\ell,\ell)$ and the
 $S$-ratio,
 $A$ is given in both the helicity and maximal bases. For the direct
 measurements
 using  $\cos(\theta_{+}) \cos(\theta_{-})$, only results for
 the basis utilized for the measurement are given. The
 uncertainties quoted are first statistical and then systematic. The
 SM predictions are $A_{\rm helicity}^{\rm SM}=0.31$ and
 $A_{\rm maximal}^{\rm SM}=0.44$.
 \label{tab:A_results}}
\begin{ruledtabular}
\begin{tabular}{c|c@{\hskip 0.3in}c|cc}
   & $\Delta \phi (\ell,\ell)$ & $S$-ratio      & $\cos(\theta_{+}) \cos(\theta_{-})_{\rm helicity}$ & $\cos(\theta_{+}) \cos(\theta_{-})_{\rm maximal}$  \\      \hline                                                                                                                                                              
   & \multicolumn{2}{c|}{indirect extraction}     & \multicolumn{2}{c}{direct extraction}  \\      \hline                                                                                                                                                              
\hline
$A_{\rm helicity}^{\rm measured}$ & $0.37 \pm 0.03\pm0.06$ & $0.27 \pm 0.03\pm0.04$ & $0.23 \pm 0.06\pm0.07$ & ---                     \\
$A_{\rm maximal}^{\rm measured}$  & $0.52 \pm 0.04\pm0.08$ & $0.38 \pm 0.05\pm0.06$ & ---                    & $0.36 \pm 0.06\pm0.08$  \\
 \end{tabular}
 \end{ruledtabular}
 \end{table*}
All results are summarized in Table~\ref{tab:A_results}. Within
uncertainties, all values are in agreement with 
the SM prediction and with each other.

\subsection{Single-lepton channel}

The measured value of $f_{\rm SM}$ using the simultaneous fit to the
\ddq\ and \dbq\ variables in the single-lepton channel 
is  $f_{\rm SM} = 1.12 \pm 0.11~\textrm{(stat.)} \pm 0.22~\textrm{(syst)}$.
Again, under the assumption that there is only SM $t \bar
t$ spin correlation,  vanishing $t \bar t$ spin correlation or any
mixture of both, this results in an indirect extraction of
$A_{\rm helicity}^{\rm measured} = 
0.35 \pm 0.03~\textrm{(stat.)}~\pm 0.08~\textrm{(syst)}$.
The systematic uncertainties and their effect on the measurement of
$f_{\rm SM}$ are listed in Table~\ref{tab:systematics_lj}. Part of the
detector modeling uncertainties were determined using nuisance parameters, 
corresponding to the uncertainties on lepton identification, $b$-jet 
tagging and jet energy calibration (denoted ``Detector modeling I'' in
Table~\ref{tab:systematics_lj}).
Uncertainties due to lepton reconstruction, jet reconstruction and resolution, 
and multijet background shape are evaluated using ensemble tests and are 
included in the ``Detector modeling II'' entry. In the single-lepton
channel, the main systematic uncertainty arises from parton showering and
fragmentation. The parametrization of $f_{\rm SM}$ versus the top
quark mass is  $\Delta f_{\rm SM} = 0.024 (m_t/\rm GeV  - 172.5)$.

\begin{table}[b]
\caption{Systematic uncertainties on $f_{\rm SM}$ determined from the
  simultaneous
  fit to \ddq\ and \dbq. Uncertainty on the background normalization
  is included
  in the statistical uncertainty of the fit while uncertainty on the
  background shape
  is included into ``Detector modeling I'' and ``Detector modeling
  II''.
  The detector modeling uncertainties are split into nuisance
  parameter uncertainties (I)
  and uncertainties evaluated via ensemble tests (II).
  \label{tab:systematics_lj}}
  \begin{ruledtabular}
  \begin{tabular}{cc}
  Source of uncertainty                   &     \\ 
  \hline 
  \multicolumn{2}{l}{Detector modeling} \\
  \hline
  Detector modeling I                  & \ppm 0.09   \\
  Detector modeling II                  & \ppm 0.02   \\
  \hline
  \multicolumn{2}{l}{Signal and background modelling}\\
  \hline
  Renormalization/factorization scale 		  &   \ppm0.06  \\
  Parton shower and fragmentation     		  &   \ppm0.16  \\
  ISR/FSR	                           	  &   \ppm0.07  \\
  Underlying event                    		  &   \ppm0.05  \\
  Color reconnection                 		  &   \ppm0.01  \\
  PDF uncertainty                         	  &   \ppm0.02  \\
  MC statistics                           	  &   \ppm0.05  \\
  Top $p_{\rm T}$ reweighting                     &   \ppm0.02 \\
  \hline                                        		     
  Total systematic uncertainty              	  &   \ppm0.22	\\
  Data statistics                         	  &   \ppm0.11     \\
  \end{tabular}
  \end{ruledtabular}
\end{table}

\begin{figure*}[htpb!]
\begin{center}
\subfigure[]{\includegraphics[width=0.49\textwidth]{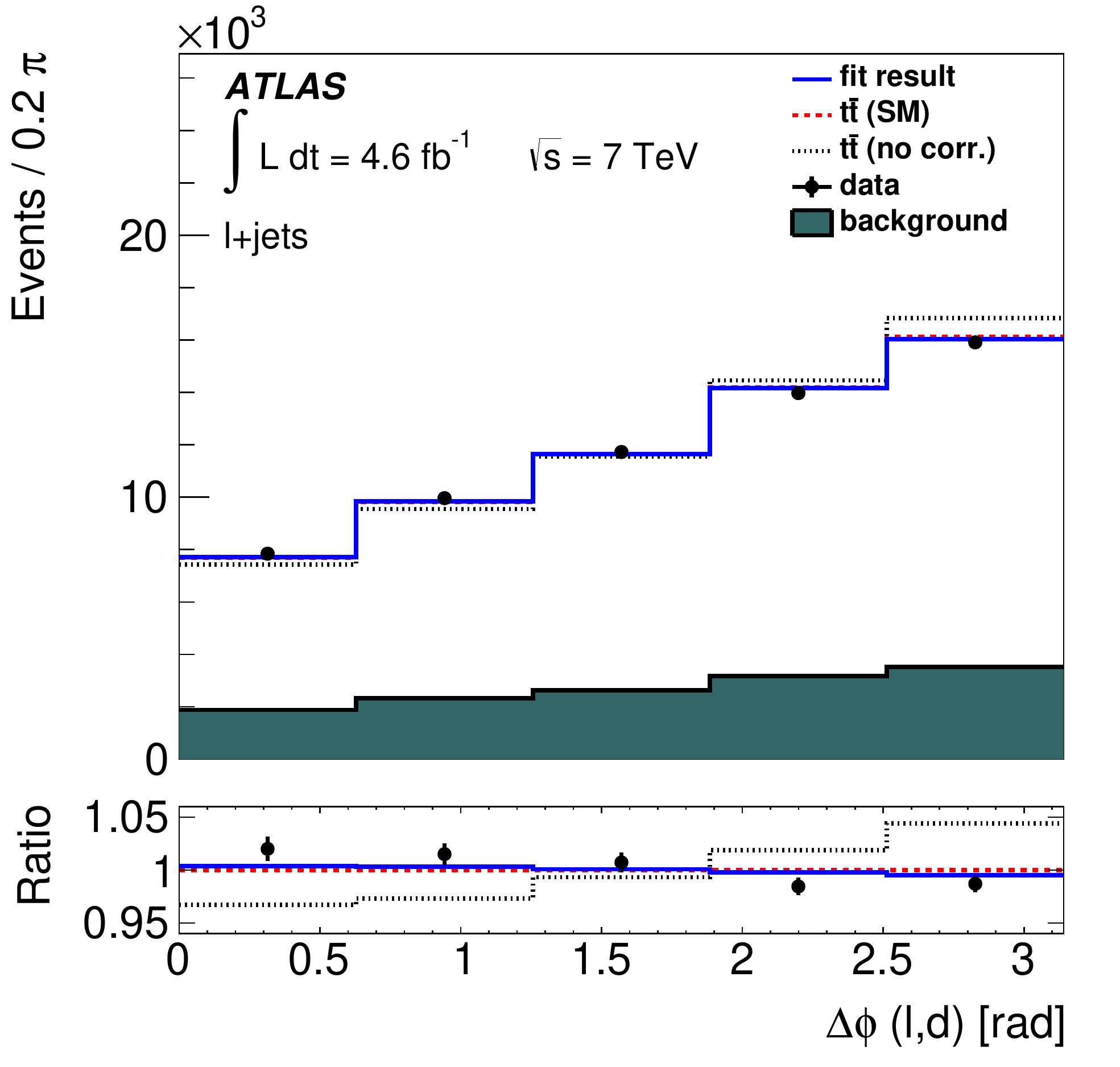}}
\subfigure[]{\includegraphics[width=0.49\textwidth]{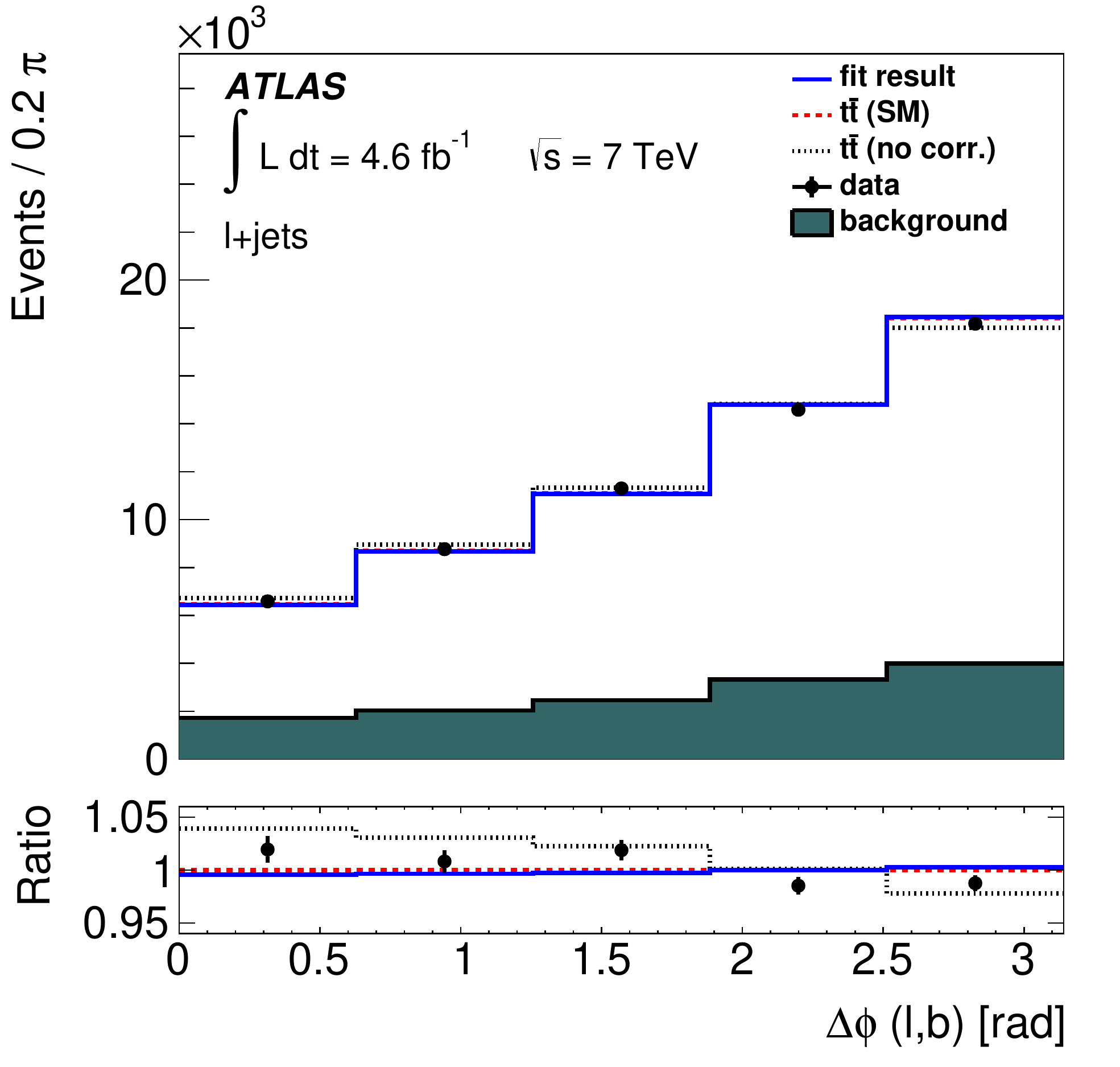}}
\end{center}
\caption{Distributions of (a) $\Delta\phi (\ell,d)$ and 
(b) $\Delta\phi (\ell,b)$ in the single-lepton final sate. The result
  of the fit to data (blue lines) is compared 
to the templates for background plus \ttbar\ signal with SM spin
correlation (red dashed lines) and without spin correlation (black
dotted lines). The bottom panel shows the ratio of the data (black
points), of the best fit (blue solid lines) and of the no spin prediction to
the SM prediction.   
}
\label{fig:finalplotslj}
\end{figure*}

Figure~\ref{fig:finalplotslj} shows the
observables including the result of the fit to data.
\begin{figure*}[htpb!]
\begin{center}
{\includegraphics[width=0.78\textwidth]{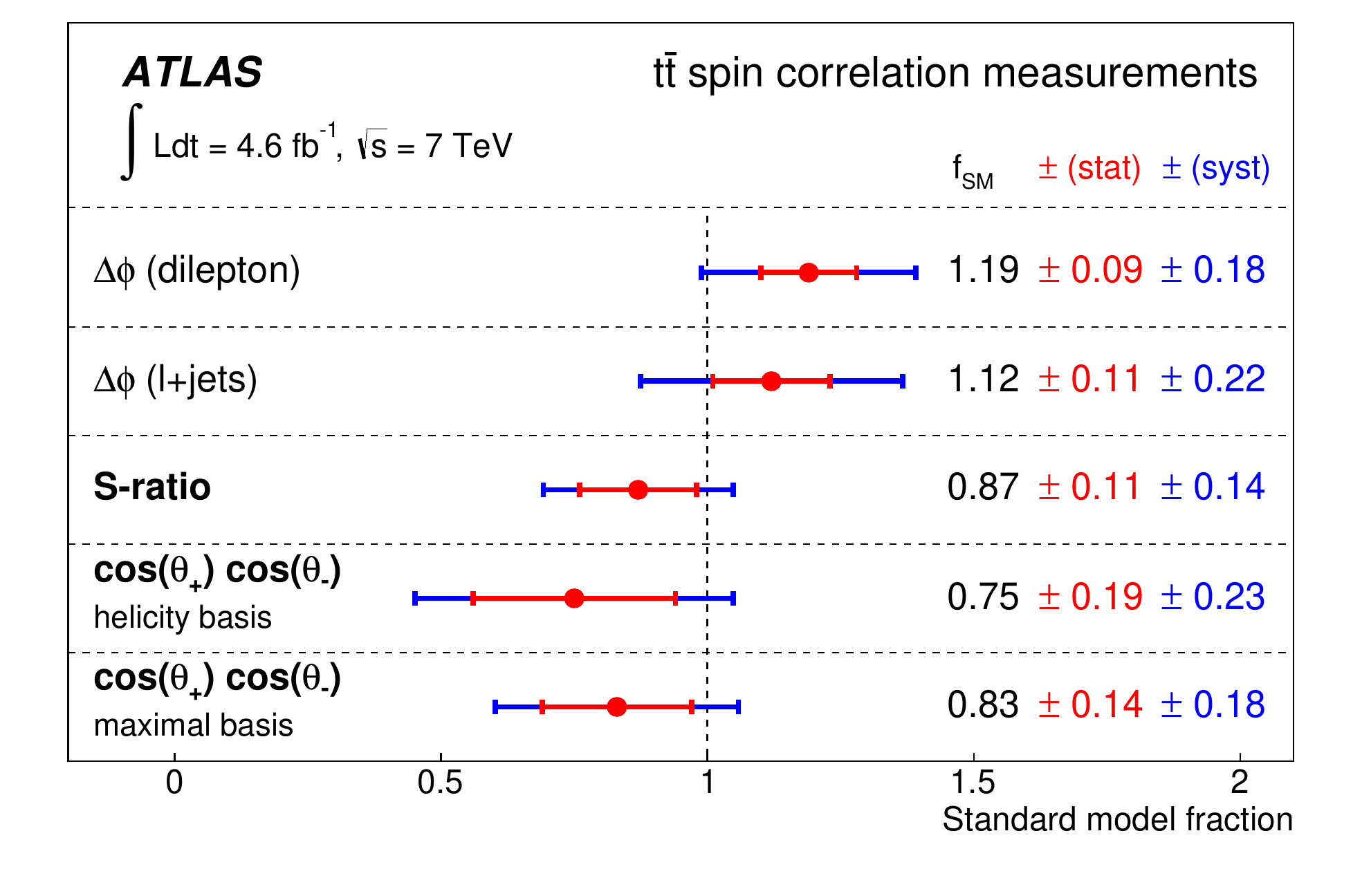}}
\end{center}
\caption{Summary of the measurements of the fraction of $t\bar{t}$
  events corresponding to the SM spin
  correlation hypothesis, $f_{\rm SM}$, in the 
  dilepton final state, using four spin correlation observables 
  sensitive to different properties of the production mechanism, and in 
  the single-lepton final state. Dashed vertical line at $f_{\rm SM} = 1$
  indicates the SM prediction. The
  inner, red error bars indicate statistical uncertainties, the outer, 
  blue error
  bars indicate the contribution of the systematic uncertainties to
  the total uncertainties.}
\label{fig:summary}
\end{figure*}

Figure~\ref{fig:summary} 
summarizes the $f_{\rm SM}$ values measured using various observables 
in the dilepton and single-lepton final states. 
All measurements agree with the SM prediction of $f_{\rm SM}=1$. 

\section{Conclusions}
\label{sec:conclusion}
The $t\bar{t}$ spin correlation in dilepton and single-lepton final states
is measured utilizing ATLAS data, corresponding to an integrated 
luminosity of $4.6$~fb$^{-1}$, recorded in proton--proton scattering at the LHC 
at a center-of-mass energy of 7~\tev. 

In dilepton final states, four observables are used with
different sensitivities to like-helicity  gluon--gluon initial states
and unlike-helicity gluon--gluon or $q\bar{q}$ initial states.
For the first time, the measurement of \ttbar\ spin correlation is performed 
using the $S$-ratio.
Also, a direct measurement of the spin correlation strengths $A_{\rm
  helicity}$ and $A_{\rm maximal}$ is performed using $\cos\theta_{+}
\cos\theta_{-}$ in the helicity and maximal bases, respectively. 
The measurement in the maximal basis is performed for the first time resulting 
in $A_{\rm maximal}^{\rm measured}= 0.36 \pm 0.10~({\rm stat} \oplus {\rm syst})$.

In the dilepton channel, the measurement of \ttbar\ spin
correlation using the azimuthal angle between the charged leptons, 
$\Delta \phi$, gives $f_{\rm SM}=1.19 \pm 0.20~({\rm stat} \oplus {\rm
  syst})$.
In the single-lepton channel, the \ttbar\ spin correlation strength
is measured for the first time at the LHC using a  
simultaneous fit to the azimuthal angle between charged lepton and $d$-quark 
\ddq\ and between charged lepton and $b$-quark \dbq. The result is 
$f_{\rm SM}=1.12\pm 0.24~({\rm stat} \oplus {\rm syst})$.
These measurements
in the dilepton and 
single-lepton channels are in good agreement with the SM predictions.

\clearpage

\section*{Acknowledgements}

We thank CERN for the very successful operation of the LHC, as well as the
support staff from our institutions without whom ATLAS could not be
operated efficiently.

We acknowledge the support of ANPCyT, Argentina; YerPhI, Armenia; ARC,
Australia; BMWFW and FWF, Austria; ANAS, Azerbaijan; SSTC, Belarus; CNPq and FAPESP,
Brazil; NSERC, NRC and CFI, Canada; CERN; CONICYT, Chile; CAS, MOST and NSFC,
China; COLCIENCIAS, Colombia; MSMT CR, MPO CR and VSC CR, Czech Republic;
DNRF, DNSRC and Lundbeck Foundation, Denmark; EPLANET, ERC and NSRF,
European Union;  
IN2P3-CNRS, CEA-DSM/IRFU, France; GNSF, Georgia; BMBF, DFG, HGF, MPG and AvH
Foundation, Germany; GSRT and NSRF, Greece; ISF, MINERVA, GIF, I-CORE
and Benoziyo Center, 
Israel; INFN, Italy; MEXT and JSPS, Japan; CNRST, Morocco; FOM and NWO,
Netherlands; BRF and RCN, Norway; MNiSW and NCN, Poland; GRICES and
FCT, Portugal; MNE/IFA, Romania; MES of Russia and ROSATOM, Russian
Federation; JINR; MSTD, 
Serbia; MSSR, Slovakia; ARRS and MIZ\v{S}, Slovenia; DST/NRF, South Africa;
MINECO, Spain; SRC and Wallenberg Foundation, Sweden; SER, SNSF and Cantons of
Bern and Geneva, Switzerland; NSC, Taiwan; TAEK, Turkey; STFC, the Royal
Society and Leverhulme Trust, United Kingdom; DOE and NSF, United States of
America.

The crucial computing support from all WLCG partners is acknowledged
gratefully, in particular from CERN and the ATLAS Tier-1 facilities at
TRIUMF (Canada), NDGF (Denmark, Norway, Sweden), CC-IN2P3 (France),
KIT/GridKA (Germany), INFN-CNAF (Italy), NL-T1 (Netherlands), PIC (Spain),
ASGC (Taiwan), RAL (UK) and BNL (USA) and in the Tier-2 facilities
worldwide.

\onecolumngrid 
\clearpage
\input{atlas_authlist}

\end{document}

%% file: atlas_authlist.tex
\begin{flushleft}
{\Large The ATLAS Collaboration}

\bigskip

G.~Aad$^{\rm 84}$,
B.~Abbott$^{\rm 112}$,
J.~Abdallah$^{\rm 152}$,
S.~Abdel~Khalek$^{\rm 116}$,
O.~Abdinov$^{\rm 11}$,
R.~Aben$^{\rm 106}$,
B.~Abi$^{\rm 113}$,
M.~Abolins$^{\rm 89}$,
O.S.~AbouZeid$^{\rm 159}$,
H.~Abramowicz$^{\rm 154}$
% ...
\bigskip
\\
% ...
$^{11}$ Institute of Physics, Azerbaijan Academy of Sciences, Baku, Azerbaijan\\
% ...
$^{84}$ CPPM, Aix-Marseille Universit{\'e} and CNRS/IN2P3, Marseille, France\\
% ...
$^{89}$ Department of Physics and Astronomy, Michigan State University, East Lansing MI, United States of America\\
% ...
$^{106}$ Nikhef National Institute for Subatomic Physics and University of Amsterdam, Amsterdam, Netherlands\\
% ...
$^{112}$ Homer L. Dodge Department of Physics and Astronomy, University of Oklahoma, Norman OK, United States of America\\
$^{113}$ Department of Physics, Oklahoma State University, Stillwater OK, United States of America\\
% ...
$^{116}$ LAL, Universit{\'e} Paris-Sud and CNRS/IN2P3, Orsay, France\\
% ...
$^{152}$ Institute of Physics, Academia Sinica, Taipei, Taiwan\\
% ...
$^{154}$ Raymond and Beverly Sackler School of Physics and Astronomy, Tel Aviv University, Tel Aviv, Israel\\
% ...
$^{159}$ Department of Physics, University of Toronto, Toronto ON, Canada\\
% ...
\end{flushleft}
